\documentclass[twocolumn]{aastex701}
\usepackage{graphicx}
\usepackage{natbib}
\usepackage[utf8]{inputenc}
\usepackage{CJK}
\usepackage{lmodern}

\shortauthors{Le Bourdais et al.}
\accepted{September 22, 2025}

\graphicspath{{./}{}}
\newcommand{\teff}{$T_{\rm eff}$}
\newcommand{\logg}{$\log g$}
\usepackage{CJK}
\begin{document}

\begin{CJK*}{UTF8}{gbsn}
\title{Tracing Planetary Accretion in a 3 Gyr-old Hydrogen-Rich White Dwarf: The Extremely Polluted Atmosphere of LSPM\,J0207+3331}

\correspondingauthor{\'{E}rika Le Bourdais}
\email{erika.le.bourdais@umontreal.ca}

\author[orcid=0000-0002-3307-1062,sname='Le Bourdais']{\'{E}rika Le Bourdais}
\affiliation{D\'epartment de physique, Universit\'e de Montr\'eal, Ave. Th\'er\`ese-Lavoie-Roux Montr\'eal, QC H2V 0B3, Canada}
\affiliation{Trottier Institute for Research on Exoplanets, Universit\'e de Montr\'eal, Ave. Th\'er\`ese-Lavoie-Roux Montr\'eal, QC H2V 0B3, Canada}
\affiliation{Centre de recherche en astrophysique du Qu\'ebec, Montr\'eal, QC, Canada}
\email{erika.le.bourdais@umontreal.ca}

\author[0000-0003-4609-4500]{Patrick Dufour}
\affiliation{D\'epartment de physique, Universit\'e de Montr\'eal, Ave. Th\'er\`ese-Lavoie-Roux Montr\'eal, QC H2V 0B3, Canada}
\affiliation{Trottier Institute for Research on Exoplanets, Universit\'e de Montr\'eal, Ave. Th\'er\`ese-Lavoie-Roux Montr\'eal, QC H2V 0B3, Canada}
\affiliation{Centre de recherche en astrophysique du Qu\'ebec, Montr\'eal, QC, Canada}
\email{patrick.dufour@umontreal.ca}

\author[0000-0001-9834-7579]{Carl Melis}
\affiliation{Department of Astronomy \& Astrophysics, University of California San Diego, La Jolla, CA 92093-0424, USA}
\email{cmelis@ucsd.edu}

\author[0000-0001-5854-675X]{Beth L. Klein}
\affiliation{Department of Physics and Astronomy, University of California, Los Angeles, CA 90095-1562, USA}
\email{kleinb@ucla.edu}

\author[0000-0002-3553-9474]{Laura K. Rogers}
\affiliation{Gemini Observatory/NOIRLab, 950 N. Cherry Ave, Tucson, AZ 85719, USA}
\email{laura.rogers@noirlab.edu}

\author[orcid=0000-0002-2384-1326,sname='B\'edard']{Antoine B\'edard}
\affiliation{Department of Physics, University of Warwick, Gibbet Hill Road, Coventry CV4 7AL, UK}
\email{antoine.bedard@warwick.ac.uk}  

\author[0000-0002-1783-8817]{John Debes}
\affiliation{Space Telescope Science Institute, 3700 San Martin Drive, Baltimore, MD 21218, USA}
\email{debes@stsci.edu}

\author[orcid=0009-0002-4970-3930,sname='Messier']{Ashley Messier}
\affiliation{Department of Astronomy, Smith College, Northampton, MA 01063, USA}
\email{amessier@smith.edu}

\author[0000-0001-6654-7859]{Alycia J. Weinberger}
\affiliation{Earth and Planets Laboratory, Carnegie Institution for Science, 5241 Broad Branch Rd NW, Washington, DC 20015, USA}
\email{aweinberger@carnegiescience.edu} 

\author[0000-0002-8808-4282]{Siyi Xu (许\CJKfamily{bsmi}偲\CJKfamily{gbsn}艺)} 
\affiliation{Gemini Observatory/NOIRLab, 950 N. Cherry Ave, Tucson, AZ 85719, USA}
\email{siyi.xu@noirlab.edu}
\begin{abstract}
We report the detection of 13 heavy elements (Na, Mg, Al, Si, Ca, Ti, Cr, Mn, Fe, Co, Ni, Cu, and Sr) in the photosphere of LSPM\,J0207+3331, a $\sim$3 Gyr-old hydrogen-rich white dwarf with an effective temperature comparable to that of the Sun. Upper limits on carbon, obtained through the absence of molecular CH, suggest accretion from a carbon-volatile-depleted source. 
The accreted parent body exhibits slight deficits of Mg and Si relative to Fe but otherwise bulk Earth-like abundance patterns; a reasonable interpretation is that LSPM\,J0207+3331 is accreting a massive differentiated rocky body that had a core mass fraction higher than the Earth's.
The high level of pollution indicates that substantial accretion events can still occur even after 3 Gyr of cooling. We also detect weak \ion{Ca}{2} H \& K line-core emission, making this only the second known isolated polluted white dwarf to exhibit this phenomenon and suggesting the presence of additional physical processes in or above the upper atmosphere. Our analysis also highlights the critical importance of including heavy elements in the model atmosphere structure calculations for highly polluted hydrogen-rich white dwarfs. Neglecting their contribution significantly impacts the inferred thermodynamic structure, leading to inaccuracies in derived stellar parameters. Finally, we show that the observed 11.3 \micron\ infrared excess can be explained by a single silicate dust disk rather than a two-ring disk model.
\end{abstract}
\keywords{White dwarf stars (1799), Stellar abundances (1577), Stellar atmospheres (1584), Exoplanet systems (484)}

\section{Introduction}

One of the greatest challenges in exoplanet science is determining the bulk composition and internal structure of extrasolar worlds. While we can measure exoplanet masses, radii, and atmospheric properties, their interiors--rocky cores, metallic structure, and overall elemental abundances that influence their atmospheric composition--remain inaccessible to direct observation \citep[e.g.,][]{seager_mass-radius_2007,dorn_can_2015}. White dwarf stars offer a unique and powerful solution to this limitation. When planetary debris (e.g., planetesimals, asteroids, comets, moons, or planets) ventures too close to these dense stellar remnants, it is gravitationally shredded and accreted \citep{debes_are_2002,jura_tidally_2003}, leaving behind a detailed chemical fingerprint in the star's once-pristine atmosphere of hydrogen and/or helium \citep{jura_extrasolar_2014,veras_planetary_2021, xu_chemistry_2024}. These polluted white dwarfs represent one of the only current methods for directly measuring the bulk elemental composition of extrasolar planetary material. For more than two decades now, they have unveiled a continuously growing picture of the bulk composition of evolved planetary systems, revealing refractory compositions consistent with rocky material \citep[e.g.,][]{zuckerman_chemical_2007,klein_chemical_2010,doyle_oxygen_2019} and volatile compositions ranging from Kuiper-Belt analogs \citep{xu_chemical_2017} to water rich \citep{farihi_evidence_2013, klein_discovery_2021, brouwers_asynchronous_2023b, rogers_seven_2024, trierweiler25}. This picture of accretion coming from planetary bodies is further supported by infrared excesses in the spectral energy distribution of white dwarfs \citep[e.g.,][]{zuckerman_excess_1987,graham_infrared_1990,becklin_dusty_2005,farihi_circumstellar_2016,lai_infrared_2021}, mid-infrared spectra of dusty debris disks \citep{reach_dust_2009,swan_first_2024,farihi_subtle_2025,rogers25}, gaseous emission and absorption from circumstellar disks \citep{gansicke_gaseous_2006, xu__evidence_2016, dennihy_five_2020, melis20, manser_frequency_2020, xu_modeling_2024, le_bourdais_revisiting_2024}, transiting debris \citep[e.g.,][]{vanderburg_disintegrating_2015,vanderbosch_white_2020,guidry_i_2021} and x-ray emission from ongoing accretion onto the white dwarf's atmosphere \citep{cunningham_white_2022,estrada-dorado_xmm-newton_2023}.

However, the physics of white dwarf atmospheres -- specifically, stellar parameters such as the star's effective temperature and main constituent (H or He) -- strongly impacts which systems can provide detailed compositional information. Past analyses have been biased toward cool (\teff $<$ 20,000 K), helium-rich white dwarfs for three key physical reasons:
\begin{enumerate}
    \item This temperature regime produces strong optical transitions for the most important rock-forming elements: oxygen, iron, magnesium and silicon. Observations of hotter white dwarfs are generally
    less likely to provide accurate abundances as radiative levitation becomes important and could artificially hold intrinsic heavy elements on top of the photosphere and bias the abundance analysis \citep[e.g., ][]{chayer_radiative_1995,chayer_improved_1995,koester_frequency_2014,ould_rouis_constraints_2024},
    \item He-atmospheres are significantly more transparent than H-atmospheres, meaning that less metal pollution is needed for a spectroscopic detection; and
    \item For white dwarfs with \teff $> 10,000$\,K, heavy elements diffuse through He-atmospheres much more slowly, with sinking timescales of hundreds of thousands to millions of years compared to days or centuries in hydrogen-rich white dwarfs. For temperatures cooler than 10,000\,K, the sinking timescales for He-atmospheres are still longer than for H-atmospheres, but the difference between the two is smaller ($\sim$2 orders of magnitude instead of about 7.)
\end{enumerate}

The last two reasons are the main factors playing into our expectation of finding little to no pollution in the atmosphere of cool hydrogen-rich white dwarfs. This physical parameter-driven selection has created a significant observational bias in our understanding of planetary compositions. Of the 2300+ known polluted white dwarfs with determined abundances or upper limits of Ca according to the Montreal White Dwarf Database\footnote{\url{https://www.montrealwhitedwarfdatabase.org/}}  \citep{dufour_montreal_2017}, the systems with the richest elemental inventories are exclusively from He-atmospheres. The current record holders--GD~362 and WD~1145+017 with 16 detected elements each \citep{zuckerman_chemical_2007,xu_two_2013,le_bourdais_revisiting_2024}--exemplify this trend. For hydrogen-rich white dwarfs fewer than 10 elements are typically observed, with the rare exception of the warm ({\teff}$=21,200$\,K) WD~1929+011 with 11 elements detected, but only 7 were detected from optical spectroscopy \citep{melis_accretion_2011,vennes_pressure_2011,gansicke_chemical_2012}.

This observational bias means we are missing compositional information from a potentially vast and uniquely important population of planetary systems. Hydrogen-rich white dwarfs represent the overwhelming majority of white dwarf stars, and the coolest systems among them are some of the oldest stars in our Galaxy. Still, we paradoxically detect far more polluted helium-rich white dwarfs due to their more transparent atmospheres. The combination of less transparent hydrogen atmospheres and the expectation that rich elemental inventories would be undetectable has meant that cool hydrogen-rich systems have been largely ignored for detailed compositional studies. Even when heavy elements are present, calcium is significantly harder to detect in hydrogen-rich atmospheres than in helium-rich ones for a given abundance \citep{blouin_no_2022}, reinforcing the assumption that multi-element detection would be rare. Nonetheless, cool, hydrogen-rich white dwarfs exhibiting atmospheric pollution, though rarely studied, offer a valuable opportunity to investigate the late stages of planetary system evolution beyond the main sequence.

In this work we report the discovery of the first cool hydrogen-rich white dwarf polluted by 10 or more heavy elements. We give a brief introduction to our target along with a description of our methods for fitting both the physical parameters and abundances in Section \ref{sec:obs_methods} and we discuss the impact of metal pollution on the derived parameters of cool hydrogen-rich white dwarfs using the photometric method in Section \ref{sec:impact_structure}. Results are presented in Section \ref{sec:abundances} and the nature and origin of the infrared excess is discussed in Section \ref{sec:ir-excess}. Finally, the key findings and outlook of this work are summarized in Section \ref{sec:conclusion}. 

\section{Observations and Methods}\label{sec:obs_methods}
\subsection{LSPM\,J0207+3331}\label{sec:lspm_intro}
LSPM\,J0207+3331 (Gaia DR3 325899163483416704) was first discovered by the Backyard Worlds:\,Planet 9 citizen science project and reported in \citet{debes_3_2019} as the first white dwarf with {\teff} $<$ 7000 K showing an infrared-excess coming from a dusty disk. The presence of a weak Paschen $\beta$ line initially led to its classification as a DA white dwarf. One particularity highlighted in the discovery paper is the exceptionally bright infrared excess, almost exceeding the flux of the white dwarf in the mid-infrared. After trying to fit the excess with a companion (brown dwarf or planet), \citet{debes_3_2019} concluded that a dusty disk was most likely even though the $W3$ 11.3 {\micron} excess could only properly be accounted for with a two-ring model. An updated discussion on the matter is provided in Section \ref{sec:ir-excess}. From the disk model, the estimated accretion rate of $\sim 3\times 10^7$g s$^{-1}$ made LSPM\,J0207+3331 a promising candidate to look for metal pollution. After confirming the presence of atmospheric metals through observations during our Lick Observatory Kast polluted white dwarf survey (\citealt{melis2018}; \citealt{doyle_new_2023}), we proceeded with obtaining high resolution spectroscopy.
 
\subsection{Observations}
\label{sec:observations}

Lick Observatory reconnaissance spectroscopy of LSPM\,J0207+3331 was obtained on UT 12~September~2019 with the Kast Double Spectrograph mounted on the Shane 3\,m telescope. Kast observations simultaneously obtained blue and red spectra on separate cameras; light is split between the two channels by the d57 dichroic around 5700\,\AA. After splitting, blue light was passed through the 600/4310 grism while red light was passed through the 830/8460 grating. A slit width of 1.0$'$$'$ was used resulting in resolving powers of roughly 1300 in the blue and 3200 in the red. Blue spectra covered 3430-5490\,\AA\ while red spectra covered 6430-8795\,\AA ; a total of 2~hours of on-source integration time was obtained with each camera resulting in an average signal-to-noise ratio per pixel (SNR) of $\sim$20 in the blue and $\sim$50 in the red. Results from Kast data are largely superseded by (and in agreement with) those from MagE and HIRES data, thus we do not discuss them further.

LSPM\,J0207+3331 was then observed on UT 10~November~2019 with the MagE echellette spectrograph on the Magellan Baade Telescope. Conditions were good with clear skies, but the observations occurred only a couple days before full Moon and thus with high sky brightness and at fairly high airmass of 2.2-2.3 for this northern source. The resulting seeing was 0.8 - 1''. We used the 0.5" slit and no CCD binning, which produces a spectral resolution of $\approx$8200.
Due to low blue flux from the target star and fringing in the red, the usable wavelength range is 3900-8900\,\AA . Wavelength calibration was done via ThAr lamps taken just before and after the observations. Two exposures on the target were taken of 2400\,s each resulting in final SNR near 5100\,\AA\ of 30 and near 6700\,\AA\ of 44. The data were reduced using the standard CarPy MagE pipeline \citep{kelson_evolution_2000,kelson_optimal_2003}, which flat-fields, extracts, and wavelength calibrates the spectra.

Our highest resolution observations of LSPM\,J0207+3331 were obtained on UT 08~October~2020 with the High Resolution Echelle Spectrograph \citep[HIRES,][]{vogt_hires_1994} on the Keck I telescope at Maunakea Observatory. Two exposures of 3600\,s each were taken with the blue collimator. The raw data were reduced and processed following the procedures described in \citet{klein_rocky_2011} using the {\sf MAKEE} and {\sf IRAF} softwares; the featureless white dwarf EGGR~180 was used to remove the instrumental response function before averaging spectra and merging orders. The final combined spectrum covers 3130--5950 \AA\ and has a SNR of 12 near 3850\,\AA\ and 14 near 5100\,\AA .

\begin{deluxetable}{lc}[h]
  \tablecaption{Photometric and astrometric data. \label{tab:photometry}}
  \tablehead{\colhead{Observation} \hspace{3.5cm} & \colhead{Value}}
  \startdata
  Gaia DR3 Parallax\tablenotemark{a} (mas)        &  $22.499 \pm 0.330$  \\
  \hline
  $GALEX$ NUV           &   $21.924 \pm 0.620$  \\
  Pan-STARRS DR2 $g$    &   $17.859 \pm 0.019$  \\
  Pan-STARRS DR2 $r$    &  $17.490 \pm 0.043$  \\
  Pan-STARRS DR2 $i$    &  $17.346 \pm 0.022$  \\
  Pan-STARRS DR2 $z$    &  $17.342 \pm 0.019$  \\
  Pan-STARRS DR2 $y$    &  $17.330 \pm 0.038$  \\
  $J_{\rm 2MASS}$       &  $16.600 \pm 0.115$  \\
  $H_{\rm 2MASS}$       &  $16.334 \pm 0.225$  \\
  $Ks_{\rm 2MASS}$       &  $15.925 \pm 0.274$  \\
  $W1$       &  $15.231 \pm 0.035$  \\
  $W2$       &  $14.224 \pm 0.041$  \\
  $W3$       &  $12.194 \pm 0.292$  \\
  \enddata
$^a$ Astrometric excess noise has been added in quadrature to the DR3 quoted parallax uncertainty.
\end{deluxetable}

\subsection{Atmospheric parameters}\label{sec:methods}
\begin{figure}[h]
    \centering
    \includegraphics[width=\linewidth,trim={0.25cm 0 1.3cm 1.5cm},clip]{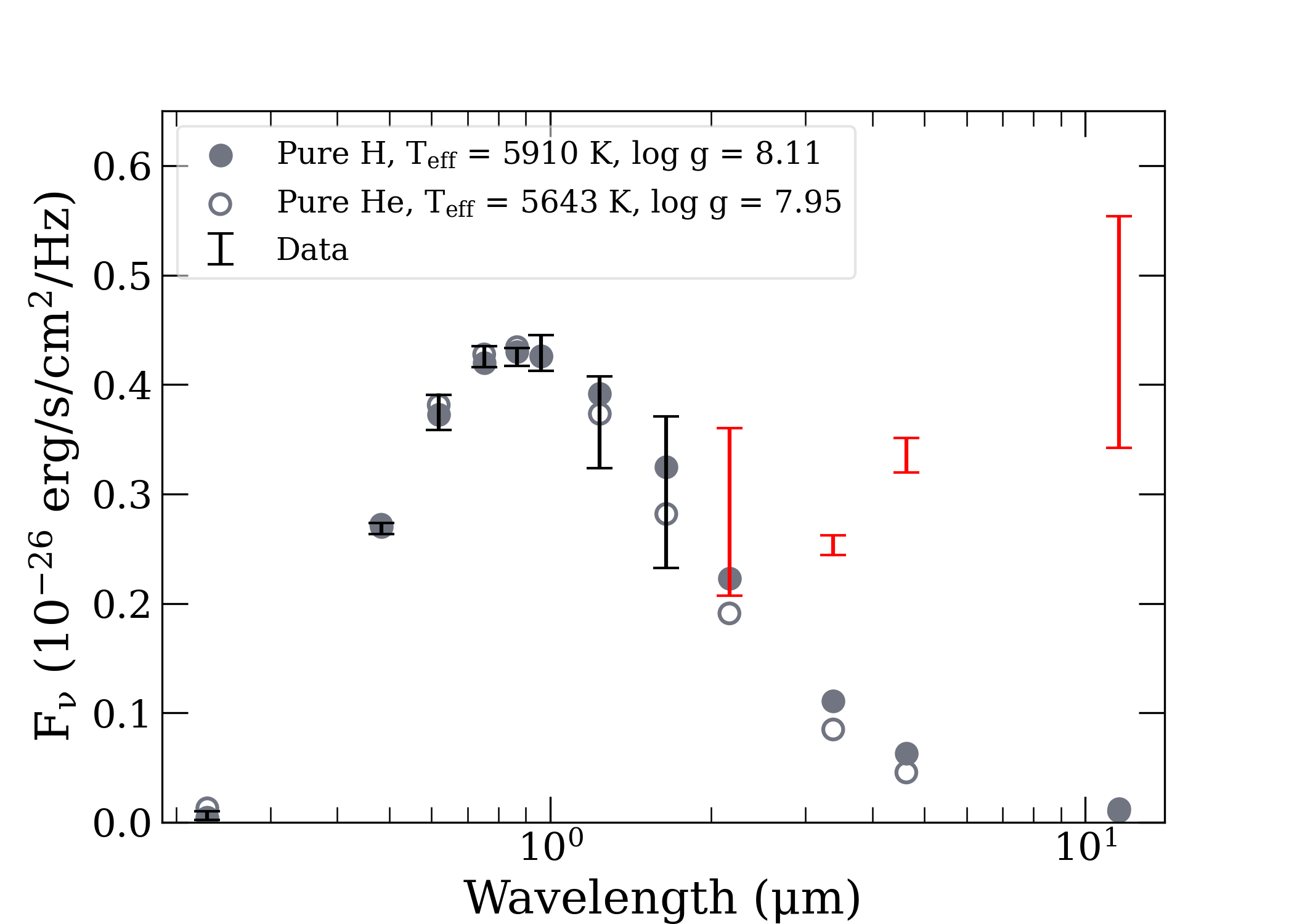}
    \caption{Photometric fit for LSPM\,J0207+3331. All spectral bands listed in Table \ref{tab:photometry} are shown with error bars. The best-fit pure hydrogen model (filled circles) is compared to a pure helium model (open circles). Photometric data excluded from the fit are highlighted in red.}
\label{fig:photometry_lspmj0207}
\end{figure}

We determined preliminary values for the effective temperature ($T_{\rm eff}$) and surface gravity ($\log g$) of this white dwarf using photometric data from GALEX NUV \citep{martin_galaxy_2005}, Pan-STARRS DR2 $grizy$ \citep{chambers_pan-starrs1_2016}, and 2MASS \citep{skrutskie_two_2006} $J$ and $H$ bands along with the \textit{Gaia} DR3 parallax \citep{gaia_collaboration_gaia_2022} (see Table~\ref{tab:photometry}). To minimize contamination from the circumstellar disk, we excluded the 2MASS $K_s$ and WISE $W1$, $W2$, and $W3$ photometric bands \citep{wright_wide-field_2010} from our analysis. Atmospheric parameters were obtained by fitting the photometric energy distribution using standard methods \citep{coutu_analysis_2019}, employing a two-dimensional grid of pure hydrogen and helium atmosphere models with $1500\,\mathrm{K} \leq T_{\rm eff} \leq 60{,}000\,\mathrm{K}$ and $7.0 \leq \log g \leq 9.0$. Fits for both pure hydrogen and helium-rich atmospheric compositions are shown in Figure~\ref{fig:photometry_lspmj0207}.

\begin{figure}
    \centering
    \includegraphics[width=\linewidth, trim={0cm 0cm 0cm 0cm},clip]{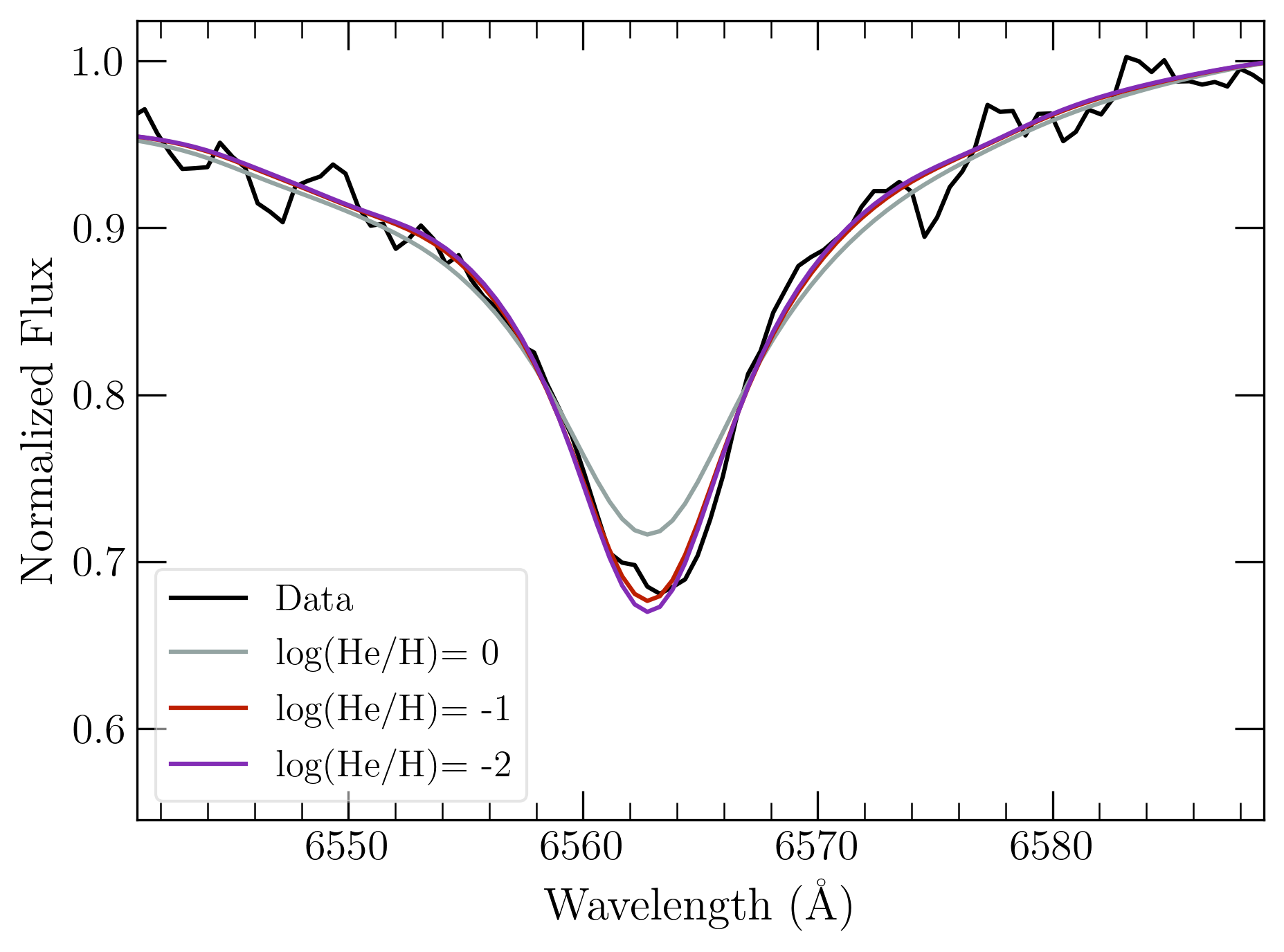}
    \caption{Models covering H$\alpha$ with He abundances of $\log({\rm He/H})= -2, -1,$ and 0 overplotted on the MagE spectrum.}
    \label{fig:fit_he}
\end{figure}

At the effective temperature of LSPM\,J0207+3331, optical photometry alone is insufficient to distinguish between hydrogen-rich and helium-rich compositions (see Figure \ref{fig:photometry_lspmj0207}). However, \citet{debes_3_2019} reported the detection of the Pa-$\beta$ line, strongly suggesting a hydrogen-rich atmosphere. This conclusion is further supported by our analysis of the optical spectrum (see Figure~\ref{fig:fit_panel01}), which reveals a highly metal-polluted white dwarf with Balmer lines, confirming that LSPM\,J0207+3331 has a hydrogen rich composition. Using the known effect of Balmer line broadening induced by the presence of He in cool hydrogen-rich white dwarfs \citep{bergeron_synthetic_1991}, we estimate an upper limit on the helium abundance of $\log ({\rm He/H}) \lesssim -1$, as larger values would noticeably affect the broadening of metal lines and the Balmer lines (see Figure \ref{fig:fit_he} and discussion below)

Adopting these atmospheric parameters, we computed a series of synthetic spectra for each heavy element detected in our spectroscopic data. We determined the abundances following the procedure outlined in \citet{dufour_detailed_2012}. In brief, for each element, we generated a dedicated grid of synthetic spectra, keeping the abundances of all other elements fixed at the values obtained from the previous iteration. Despite multiple iterations, we found that the cores of several strong absorption lines were not well reproduced. This discrepancy prompted us to investigate whether the large metal content could affect the atmospheric structure of this relatively cool white dwarf. For hydrogen-rich white dwarfs, it is typically assumed that heavy elements do not significantly alter the thermodynamic structure and are included only in the calculation of the emergent spectrum, a reasonable approximation even for strongly polluted stars \citep[see][]{gianninas_discovery_2004}. However, this assumption has not been thoroughly tested at lower effective temperatures, particularly near $T_{\rm eff} \sim 5000\,\mathrm{K}$.

\begin{figure}[h]
    \centering
    \includegraphics[width=\linewidth,trim={0.25cm 0 0.2cm 0.2cm},clip]{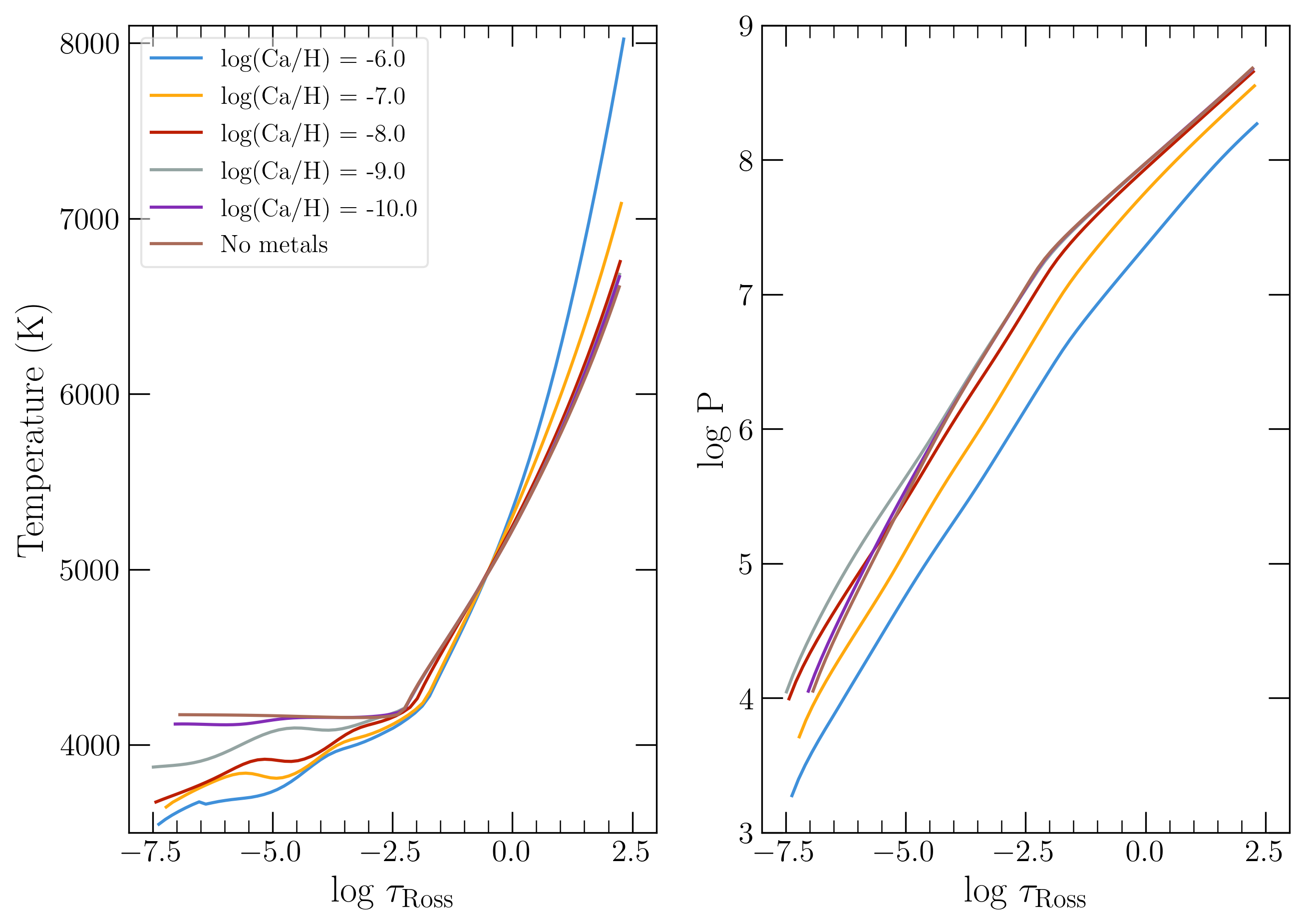}
    \caption{Temperature (left) and pressure (right) structure as a function of the optical depth for a hydrogen-rich white dwarf model with \teff\,= 5000\,K and \logg\,= 8.0. Cases explored include no metals and metal abundances between log(Ca/H) = -6 and -10.}
    \label{fig:structure_tp}
\end{figure}

In Figure~\ref{fig:structure_tp}, we present the temperature and pressure structures of a $\log g = 8.0$ model at $T_{\rm eff} = 5000\,\mathrm{K}$ with various calcium abundances (other elements scaled in chondritic proportions) explicitly included in the structural calculations. These models show that deviations from a pure hydrogen structure become significant for $\log\,({\rm Ca/H}) \gtrsim -9.0$. Motivated by this result, we recalculated the atmospheric parameters using a grid of hydrogen-rich models that self-consistently includes heavy elements in the structural calculations. An exploration of how high metal content affects the atmospheric parameters derived from photometric energy distribution of hydrogen-rich white dwarfs is presented in Section \ref{sec:impact_structure}.

Although in this particular case the inferred values of $T_{\rm eff}$ and $\log g$ do not change compared to the initial photometric fit, the synthetic spectra computed with heavy elements explicitly included in the atmospheric structure calculations yield significantly improved fits to the line cores. The adopted physical parameters derived from our best-fit hydrogen-rich + metal model are summarized in Table~\ref{tab:parameters}.

\begin{deluxetable}{lc}[h]
  \tablecaption{Physical parameters of the white dwarf LSPM\,J0207+3331. The cooling age, radius, and mass, and convection zone mass ($M_{\rm CVZ}$) are derived from our best-fit $T_{\rm eff}$ and $\log g$ values using white dwarf evolutionary models from \citet{bedard_spectral_2020}. 
  The progenitor mass and main-sequence lifetime was calculated using \textsc{wdwarfdate} \citep{kiman_wdwarfdate_2022}.\label{tab:parameters}}
  \tablehead{\colhead{Parameter} \hspace{3.5cm} & \colhead{Value}}
  \startdata
  \teff~(K)       &  $5910 \pm 98$  \\
  \logg         &   $8.11 \pm 0.03$ \\
  Main constituent & Hydrogen \\
  Spectral type & DZA \\
  Cooling age (Gyr)    &   $3.08 \pm 0.32$  \\
 Radius ($R_\odot$)    &  $0.0118 \pm 0.0004$  \\
  Mass ($M_\odot$)    &  $0.656 \pm 0.029$  \\
  $\log(M_{\rm CVZ})~~(M_{\rm WD})$       &  -6.841 \\
  Progenitor Mass ($M_\odot$) & $1.86\pm0.44$\\
  Main-sequence lifetime (Gyr) & $1.54\,^{+1.92}_{-0.56}$ \\
  \enddata
\end{deluxetable}

Still, the presence of emission in the core of \ion{Ca}{2} H \& K, features not reproduced by our photospheric models, suggests that additional physical processes may be at play in or above the upper atmospheric layers. While this emission could arise from a non-photospheric region such as a chromosphere or magnetically heated layer, it may also indicate that our current LTE models do not fully capture the thermal structure near the surface (see Section~\ref{ssec:ca_core} for further discussion).

Interestingly, the effective temperature of LSPM\,J0207+3331 is comparable to that of the Sun, and this object can be viewed as a high-gravity analog of a solar-type atmosphere, albeit with a metallicity reduced by slightly less than two orders of magnitude relative to hydrogen. A direct comparison with the high-resolution solar spectrum available from BASS2000\footnote{\url{https://bass2000.obspm.fr/download/solar_spect.pdf}} proves useful in identifying several photospheric absorption features. One striking difference, however, is the complete absence of molecular CH absorption near 4300\,\AA\  in LSPM\,J0207+3331 (see \citealt{vornanen_gj_2010} for rare cases of CH molecular features in white dwarf photospheres). While this band is among the strongest molecular features in the solar optical spectrum, its absence here is not surprising given the strongly carbon-poor composition of the accreted material. At these temperatures, atomic carbon does not produce significant optical absorption, making molecular diagnostics particularly valuable. In the case of LSPM\,J0207+3331, the non-detection of CH leads to a stringent upper limit of $\log ({\rm C/H}) < -7.3$ (see Table~\ref{tab:abundances}), significantly more restrictive than what could be achieved from the absence of C$_2$ Swan bands alone. This result provides strong evidence against the accretion of a carbon-volatile-rich parent body and highlights the utility of optical molecular diagnostics for determining carbon abundances in cool hydrogen-rich white dwarfs, particularly when ultraviolet spectroscopy is not available.

Our final abundances, derived using the improved model structure, are listed in Table~\ref{tab:abundances}. All of our abundances and our C upper limit were derived using the HIRES spectrum. The Li, O and K upper limits were derived using the MagE spectrum. Unfortunately, for oxygen, we can only derive a very weak upper limit of $\log({\rm O/H}) < -2.00$, owing to the absence of strong transitions at low effective temperatures. A detailed discussion of the individual elemental abundances and their implications is presented in Section~\ref{sec:abundances}.

\begin{deluxetable*}{ccccccc}
\tablecaption{Photospheric abundances, sinking timescales, abundance ratios relative to Fe assuming build-up phase (BU) and steady state (SS), accretion rates, and convective zone masses for LSPM\,J0207+3331. Sinking timescales are calculated assuming a pure-hydrogen background and come from \citet{fontaine_timescales_2015} and \citet{dufour_montreal_2017}, except for strontium which is from this work. Accretion rates were calculated following \citet{koester_accretion_2009}, assuming steady-state. Upper limits were excluded from the total mass and accretion rate. The lines used for fitting are listed in the Appendix.\label{tab:abundances}}
\tablehead{
\colhead{Z}&\colhead{log(Z/H)}&\colhead{$\log\tau$}&\colhead{log(Z/Si)$_{\rm BU}$}&\colhead{log(Z/Si)$_{\rm SS}$}&\colhead{$\dot{M}$}&\colhead{$M$}\\
& &(yrs)&&&(g s$^{-1}$)&(g)
}
\startdata
He & $< -1.00 $& \nodata & \nodata & \nodata & \nodata & $< 1.90\times10^{25}$\\
Li & $< -9.00 $& 5.08 & $< -2.63 $ & $ < -2.90$ & $<3.38\times10^7$ & $<1.30\times10^{18}$ \\
C  & $< -7.30$ & 5.00 & $< -0.93$ & $ < -1.12$ & $<3.54\times10^7$  &  $<1.12\times 10^{20}$\\
O  & $< -2.00$ & 5.06 & $< 4.37$ &  $< 4.13$ & $<8.30\times10^{12}$  & $<3.00\times10^{25}$\\
Na & $-7.98 \pm 0.20 $& 4.94 & $ -1.61 \pm 0.28 $ & $ -1.74 \pm 0.28 $ & $1.619\times10^7$ & $4.492\times10^{19}$\\
Mg &$ -6.41 \pm 0.20 $& 4.91 &$ -0.04 \pm 0.28 $&$ -0.14 \pm 0.28 $&  $6.911\times10^8$ &  $1.765\times10^{21}$\\
Al & $-7.28 \pm 0.10 $& 4.85 & $ -0.91 \pm 0.22 $ & $ -0.95 \pm 0.22 $ & $1.169\times10^8$ &  $2.643\times10^{20}$\\
Si & $-6.37 \pm 0.20$ & 4.81 & \nodata & \nodata & $1.090\times10^9$ &  $2.236\times10^{21}$\\
K & $< -7.50$ & 4.85 & $< -1.13 $ & $< -1.17$ & $<1.04\times10^8$ &  $<2.31\times10^{20}$\\
Ca & $-7.38 \pm 0.10$ & 4.88 & $ -1.01 \pm 0.22 $ & $ -1.08 \pm 0.22 $ & $1.308\times10^8$ &  $3.118\times10^{20}$\\
Ti & $-8.74 \pm 0.15$ & 4.83 & $ -2.37 \pm 0.25 $ & $ -2.39 \pm 0.25 $ & $7.619\times10^6$ &  $1.625\times10^{19}$\\
V & $< -9.00$ & 4.79 & $< -3.13$ & $ <-3.11$ & $<4.89\times10^6$  &  $<9.51\times10^{18}$\\
Cr & $-7.83 \pm 0.15$ & 4.78 & $ -1.46 \pm 0.25 $ & $ -1.43 \pm 0.25 $ & $7.565\times10^7$ &  $1.435\times10^{20}$\\
Mn & $-8.10 \pm 0.30$ & 4.74 & $ -1.73 \pm 0.28 $ & $ -1.66 \pm 0.28 $ & $4.642\times10^7$ &  $8.144\times10^{19}$\\
Fe & $-6.20 \pm 0.10$ & 4.73 & $0.17 \pm 0.22$ & $0.25 \pm 0.22$ & $3.853\times10^9$ &  $6.576\times10^{21}$\\
Co & $-8.44 \pm 0.20$ & 4.72 & $ -2.07 \pm 0.28 $ & $ -1.98 \pm 0.28 $ & $2.417\times10^7$ &  $3.993\times10^{19}$\\
Ni & $-7.48 \pm 0.10$ & 4.72 & $ -1.11 \pm 0.22 $ & $ -1.01 \pm 0.22 $ & $2.210\times10^8$ &  $3.627\times10^{20}$\\
Cu & $-9.21 \pm 0.25$ & 4.68 & $ -2.84 \pm 0.32 $ & $ -2.70 \pm 0.32 $ & $4.875\times10^6$ &  $7.312\times10^{18}$\\
Sr & $-10.85 \pm 0.10$ & 4.56 & $ -4.48 \pm 0.22 $ & $ -4.23 \pm 0.22 $ & $2.016\times10^5$ &  $2.310\times10^{17}$\\
\hline
\hline
Total & & & & & $6.422\times10^9$ & $1.220\times 10^{22}$
\enddata
\end{deluxetable*}

\section{Impact of metal pollution on atmospheric parameters of cool hydrogen-rich white dwarfs}
\label{sec:impact_structure}
During our spectroscopic analysis of LSPM\,J0207+3331, we found that the standard pure hydrogen atmospheric structure, when used to compute synthetic spectra including heavy elements, failed to reproduce the cores of several metal lines. This discrepancy prompted us to include heavy elements directly in the structural calculations of the atmosphere, a procedure commonly adopted for helium-rich DZ white dwarfs but not for hydrogen-rich ones. Incorporating heavy elements led to significantly improved agreement with the observed line profiles, even though the inferred values of $T_{\rm eff}$ and $\log g$ changed only slightly in this particular case.

Motivated by this result, we explored more generally how heavy metal pollution affects the determination of atmospheric parameters from photometric energy distributions, particularly in cool hydrogen-rich white dwarfs. As shown previously (Figure~\ref{fig:structure_tp}), metal polluted atmospheres can display significant changes in their temperature and pressure structures at low effective temperatures when metal content becomes high enough. Here, we examine how these structural changes propagate into synthetic photometry and how fitting such photometry using standard pure hydrogen grids may lead to systematic biases.

To this end, we constructed a three-dimensional grid of model atmospheres and synthetic spectra that include heavy elements in both the equation of state and the structural calculations. The grid spans $T_{\rm eff}$ from 4000 to 7000 K (in 250 K steps), $\log g$ values of 7.5, 8.0, and 8.5, and calcium abundances ranging from $\log(\mathrm{Ca}/\mathrm{H}) = -10$ to $-6$ (in 1 dex steps), with all other heavy elements scaled to CI chondritic abundances \citep{lodders_solar_2003} relative to calcium.

\subsection{Systematic Errors in Photometric Parameter Recovery}
To assess how these structural differences affect photometric parameter determinations, we generated artificial photometry from our synthetic spectra (assuming a distance of 10 pc) using SDSS $ugriz$, Pan-STARRS $grizy$, and 2MASS $JHK_s$ filters. We then fitted the artificial photometry using a grid of pure hydrogen models, simulating the standard approach to white dwarf parameter determination. The fitting was performed with different filter combinations: SDSS $ugriz$, Pan-STARRS $grizy$, $ugriz$+$JHK_s$, and $grizy$+$JHK_s$.

\begin{figure}[h]
    \centering
    \includegraphics[width=\linewidth]{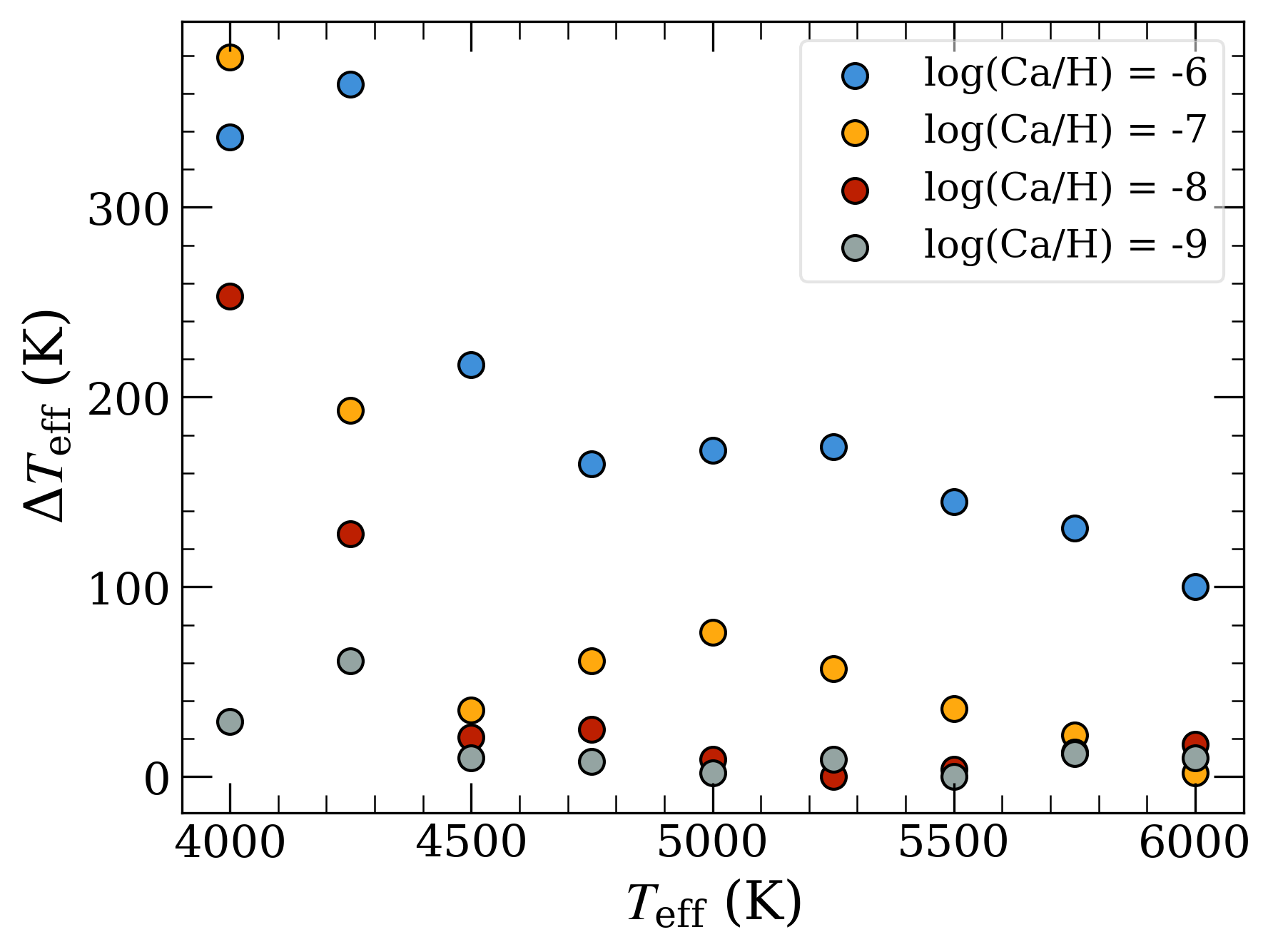}
    \includegraphics[width=\linewidth]{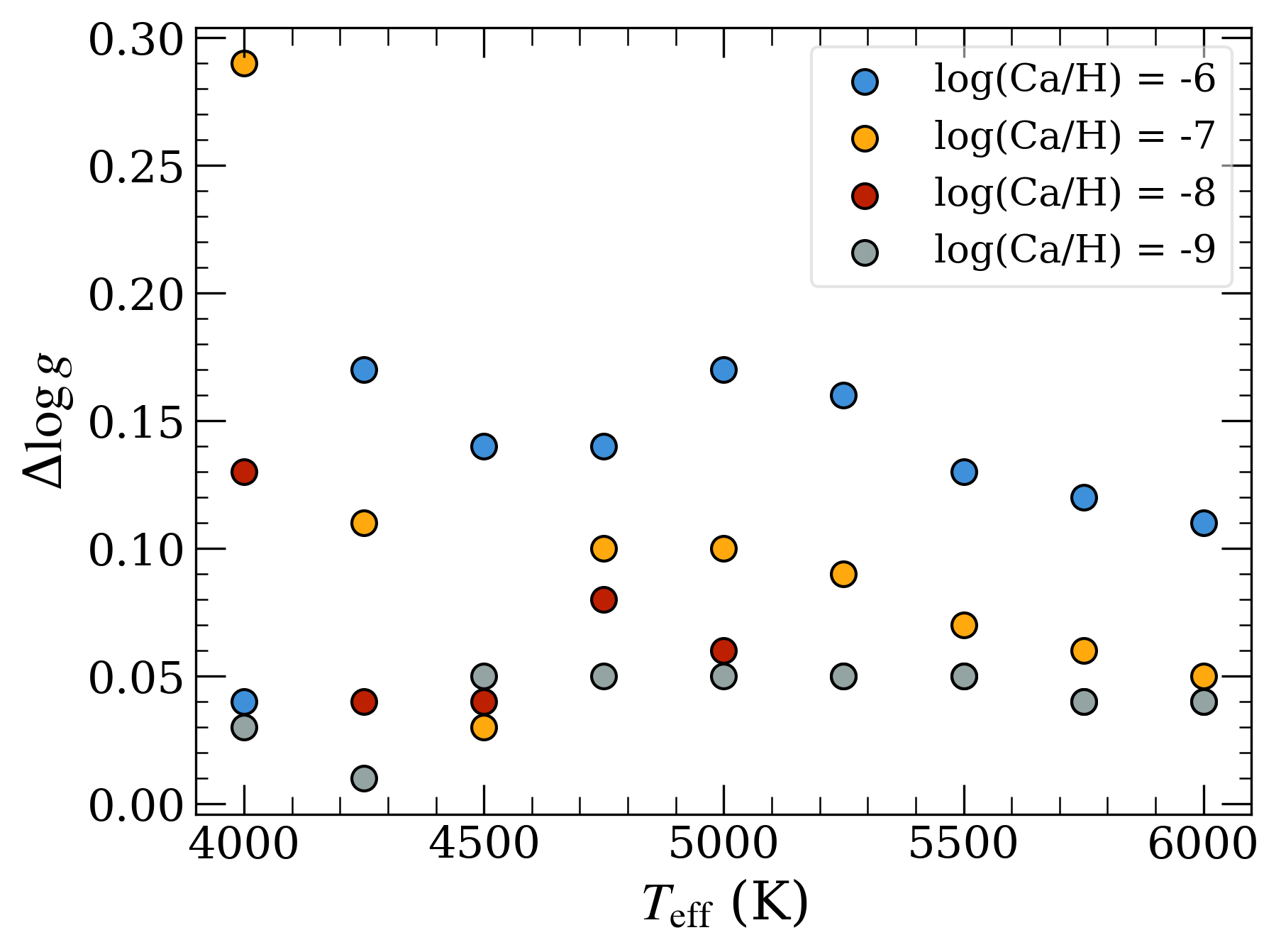}
    \caption{Deviations between fitted and the model's true {\teff} (top) and {\logg} (bottom) when models with heavy elements are fit using a pure-hydrogen grid, shown as a function of the white dwarf's true {\teff} for metal abundances from $\log({\rm Ca/H}) = -9.0$ to $-6.0$ for {\logg} = 8.0. The experiment is done using Pan-STARRS $grizy$ and 2MASS $JHK_s$ bandpasses.}
    \label{fig:delta_teff}
\end{figure}

The offsets in recovered $T_{\rm eff}$ as a function of the model's true $T_{\rm eff}$ and $\log(\mathrm{Ca}/\mathrm{H})$ are shown in Figure~\ref{fig:delta_teff}. For $\log(\mathrm{Ca}/\mathrm{H}) \gtrsim-8.0$ and $T_{\rm eff} \lesssim 5000$ K, fitting metal-polluted photometry with pure hydrogen models leads to systematic underestimations of $T_{\rm eff}$ by over 100 K and underestimations in $\log g$ of 0.1--0.2 dex, regardless of surface gravity. 

We also found that the photometric sensitivity to metal pollution depends on the filter set used. Bandpasses that sample the peak of the SED (e.g., Pan-STARRS $grizy$ + 2MASS $JHK_s$) provide better constraints than SDSS $ugriz$ alone. In particular, fitting only optical filters for polluted DAs cooler than 5000~K fails to detect structural deviations due to heavy elements, emphasizing the need to include infrared bands where possible.

\subsection{Metallicity Effects on CIA Features in the Infrared}
\begin{figure}[h]
\centering
\includegraphics[width=\linewidth,trim={0.2cm 0 0.2cm -0cm},clip]{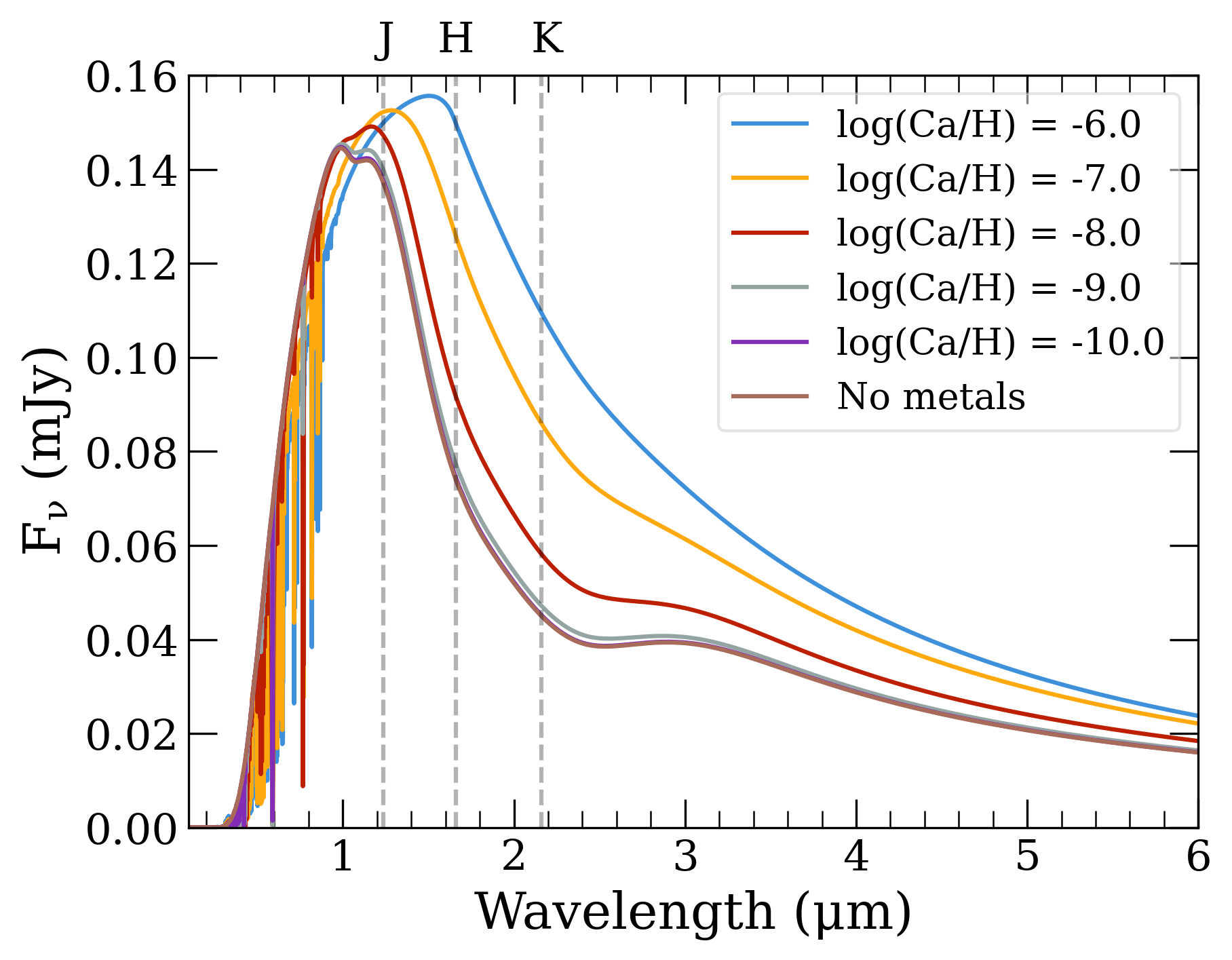}
\caption{Spectral energy distribution for a model white dwarf with {\teff} = 4000 K and {\logg} = 8.0 for various metallicities. The 2MASS $JHK_s$ spectral bands are indicated by dashed lines.}
\label{fig:SED_example}
\end{figure}

Historically, 2MASS photometry has been used cautiously in parameter determinations due to potential contamination from circumstellar dust and collision-induced absorption (CIA) by H$_2$--H$_2$. However, our calculations show that metal pollution itself modifies the CIA signature. As illustrated in Figure~\ref{fig:SED_example}, increasing metallicity reduces the strength of the CIA fundamental band near 2.5~$\mu$m by lowering the atmospheric pressure. Since CIA opacity scales with the square of density, metal-induced structural changes delay the onset of CIA absorption \citep{borysow_high-temperature_2001,dufour_spectral_2007,blouin_new_2018}. This makes $JHK_s$ photometry sensitive to both CIA and metal content, underscoring the need to account for metallicity when analyzing cool, highly polluted hydrogen-rich white dwarfs.

In summary, for hydrogen-rich white dwarfs with $T_{\rm eff} \lesssim 5000$ K and $\log(\mathrm{Ca}/\mathrm{H}) \gtrsim -8.0$, including heavy elements in model structure calculations is essential for accurate determination of atmospheric parameters from photometry. Ignoring this effect introduces systematic errors in both $T_{\rm eff}$ and $\log g$, particularly in the near-infrared.

\section{photospheric abundance analysis}\label{sec:abundances}
We detect a total of 13 metals in the photosphere of LSPM\,J0207+3331, firmly classifying it as a DZA white dwarf. The derived abundances are listed in Table~\ref{tab:abundances}, and the corresponding photospheric fits are shown in Figure~\ref{fig:fit_panel01}. 
This is only the fifth reported detection of strontium (Sr) in a white dwarf photosphere \citep{zuckerman_chemical_2007,swan_interpretation_2019,vennes_cool_2024,obrien_characterizing_2025}. Our detection suggests that Sr may be more easily probed at cooler effective temperatures. Targeted high-resolution spectroscopy of such objects may therefore prove effective in expanding the inventory of extrasolar strontium. 

\begin{figure*}
    \centering
    \includegraphics[width=\linewidth]{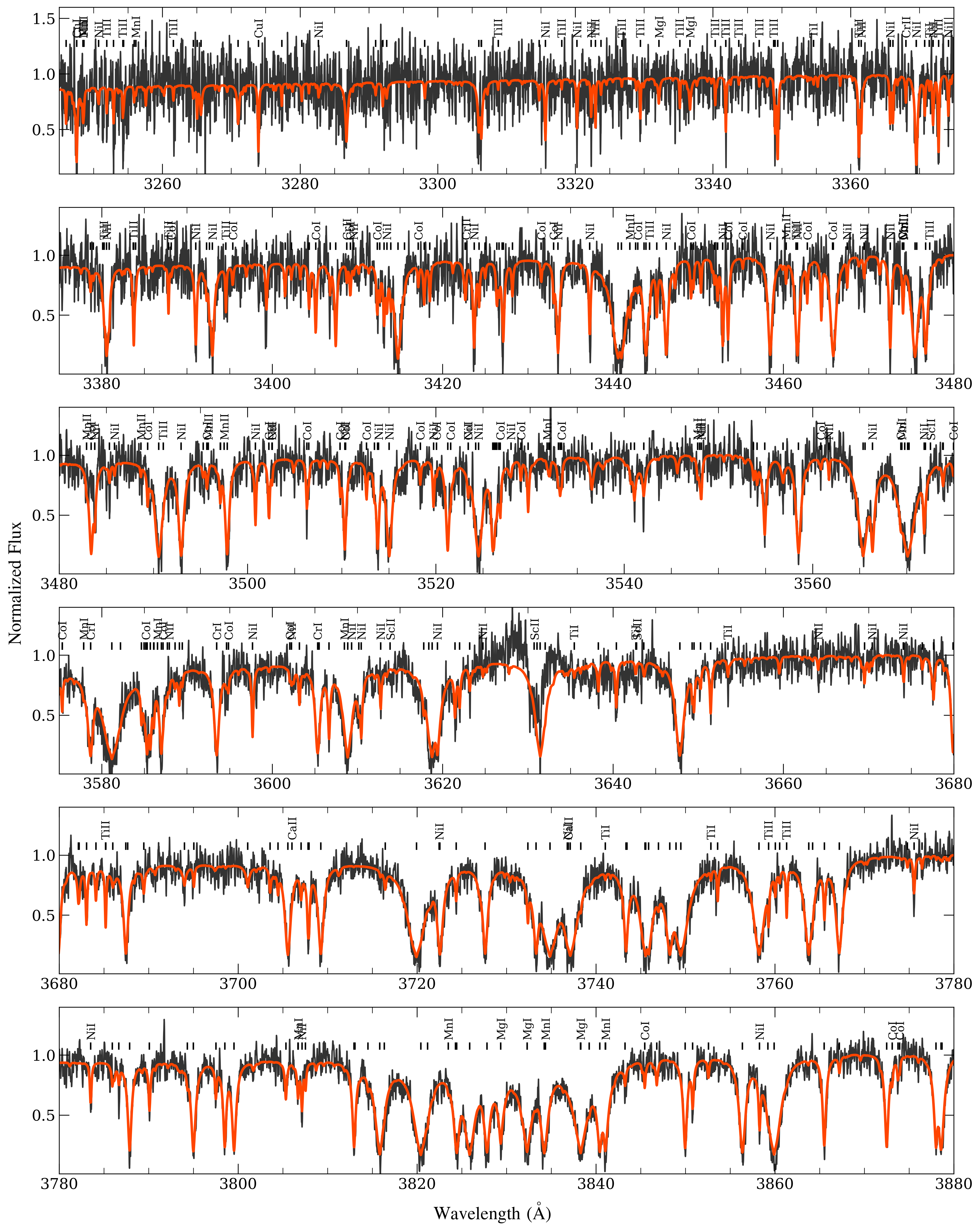}
    \caption{Best-fit model spectrum (orange) overlaid on the smoothed (five-point average window) HIRES spectrum (black) of LSPM\,J0207+3331. The feature at 3630\,\AA\ is a detector artifact. For this panel, all the unlabeled lines are Fe.}
    \label{fig:fit_panel01}
\end{figure*}

\addtocounter{figure}{-1}

\begin{figure*}
    \centering
    \includegraphics[width=\linewidth]{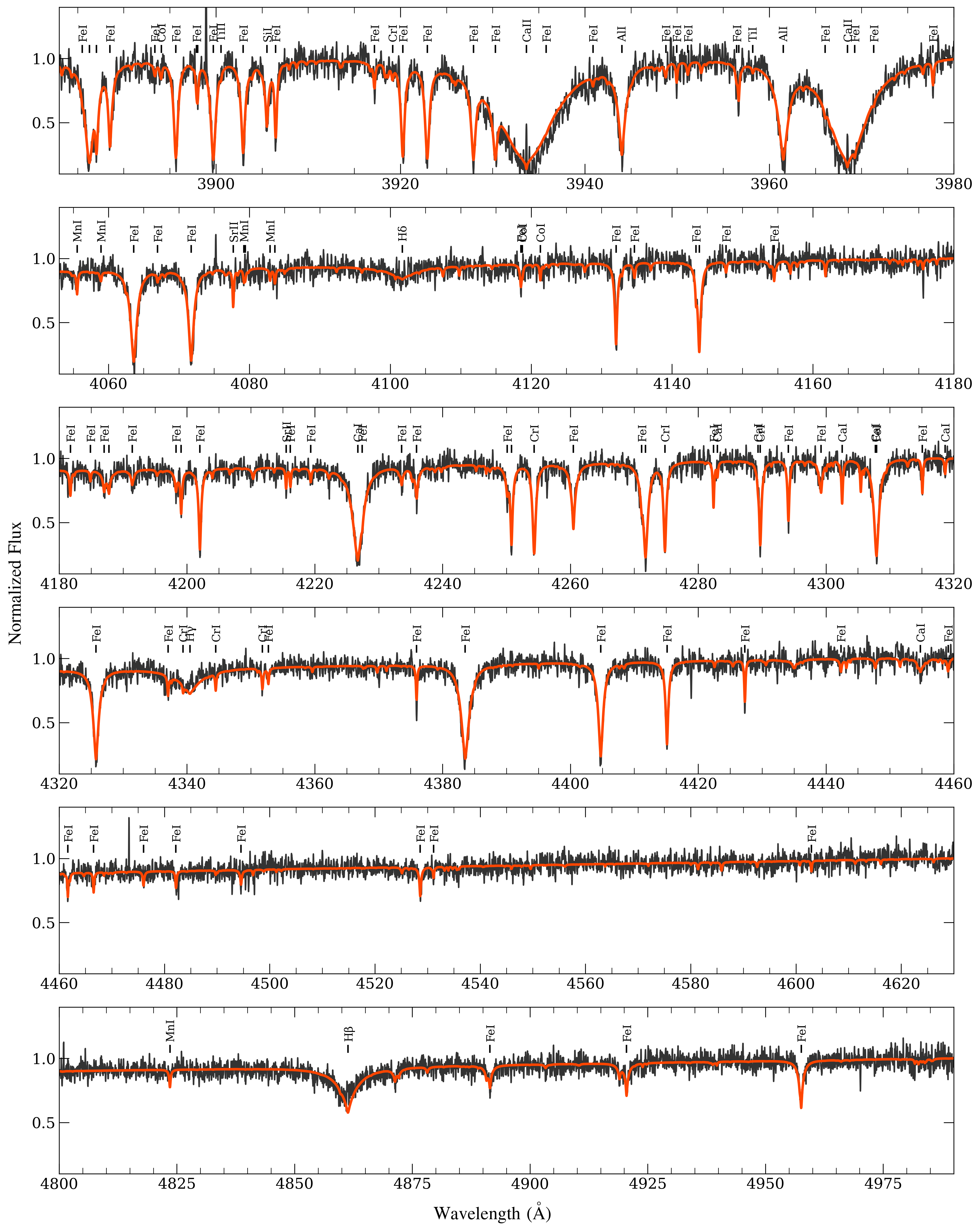}
    \caption{continued.}
    \label{fig:fit_panel02}
\end{figure*}

\addtocounter{figure}{-1}

\begin{figure*}
    \centering
    \includegraphics[width=\linewidth]{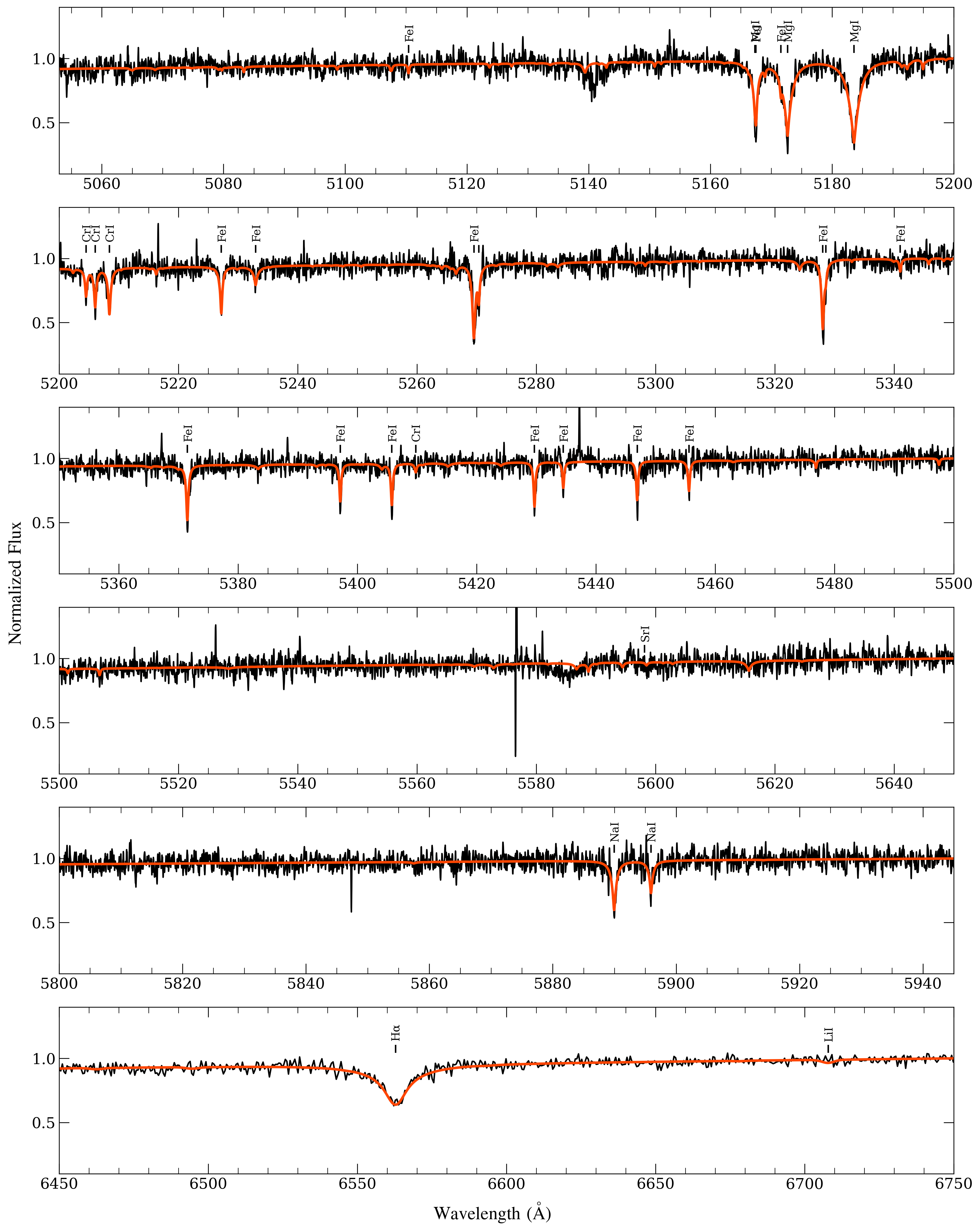}
    \caption{continued. The last panel shows the unsmoothed MagE data with our best fit model.}
    \label{fig:fit_panel03}
\end{figure*}

\begin{figure}[h]
    \centering
    \includegraphics[width=\linewidth,trim={0.2cm 0.2cm 0.2cm 0.25cm},clip]{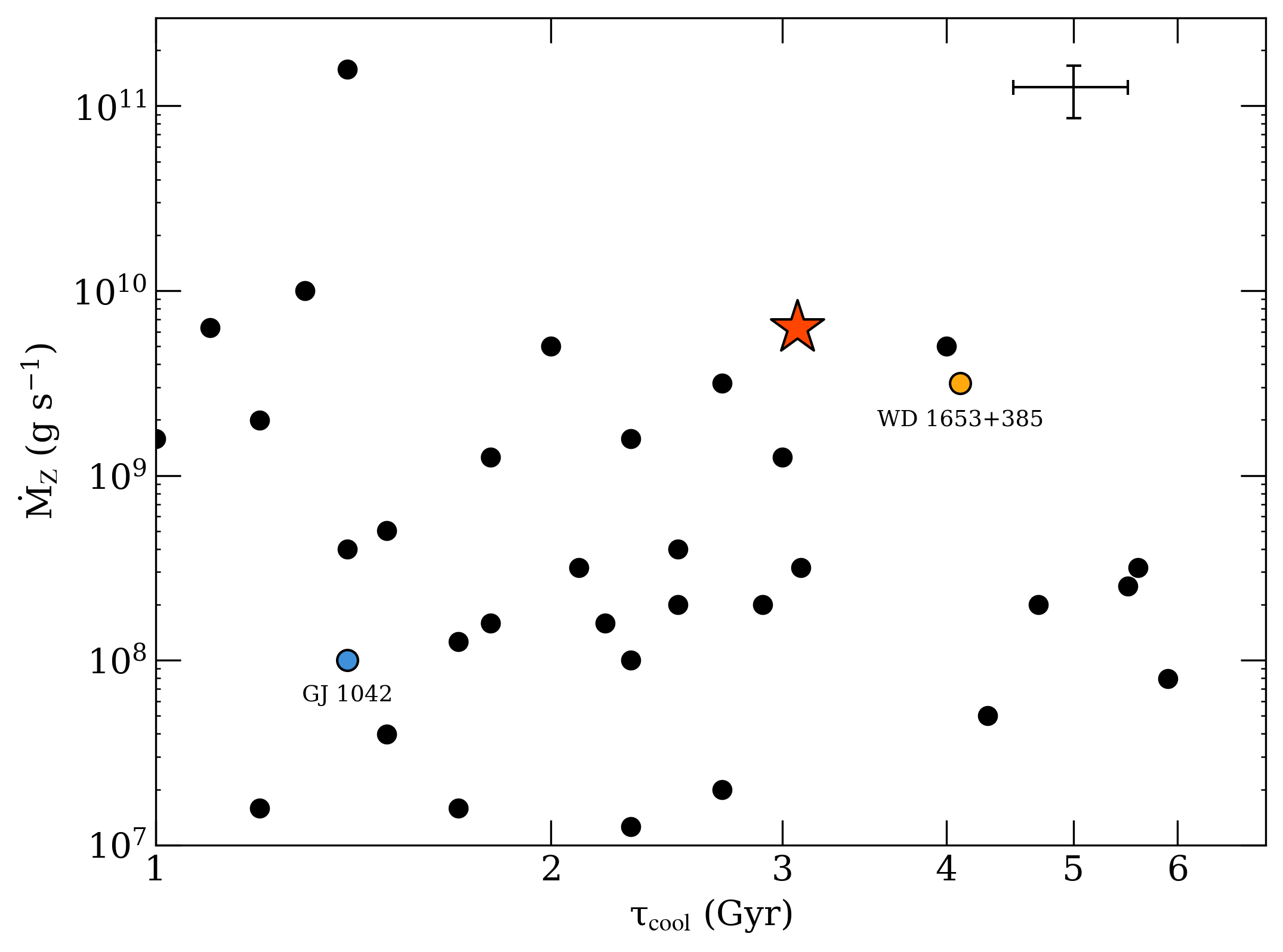}
    \includegraphics[width=\linewidth,trim={0.2cm 0.2cm 0.2cm 0.1cm},clip]{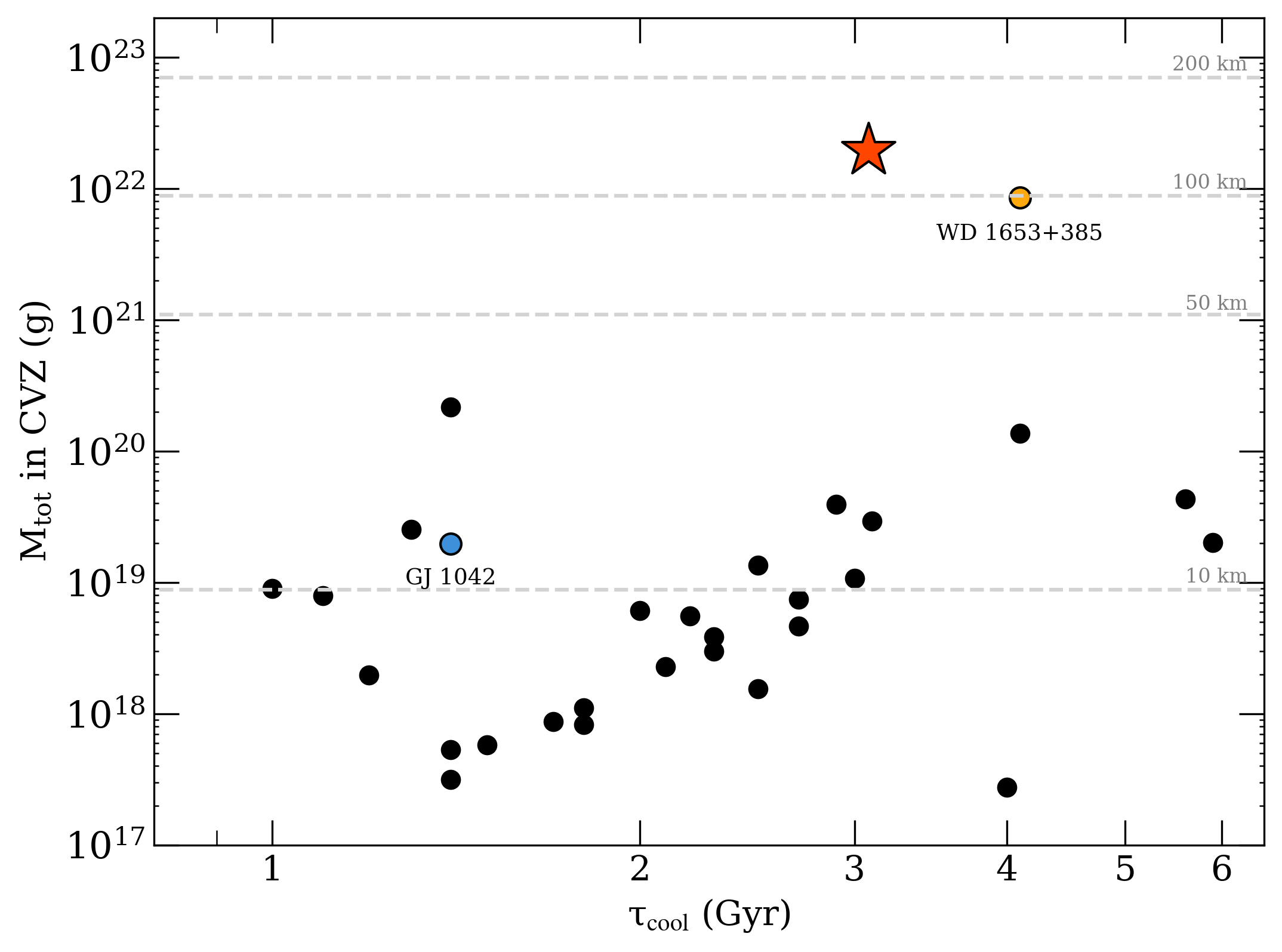}
    \caption{Top: Reproduction of Figure 3 of \citet{blouin_no_2022}, which shows the accretion rates as a function of cooling age for hydrogen dominated white dwarfs, with the accretion rate for LSPM\,J0207+3331 added (red star). All stars with only a Ca abundance are represented in black. The typical error bar is represented in the upper right corner. Bottom: Total mass of metals in the convection zone of the DAZ sample in \citet{blouin_no_2022} and LSPM J0207+3331 estimated from the Ca mass and assuming that Ca represents 1.6\% of the total mass. The horizontal lines indicate the mass corresponding to a spherical asteroid with radius of 10, 50, 100 and 200 km assuming a density of 2.1\,g\,cm$^{-3}$. Stars with detailed spectroscopic analysis are colored in blue \citep[GJ~1042, $N_{\rm Z} = 4$, ][]{zuckerman_metal_2003} and yellow \citep[NLTT~43806 $=$ WD~1653+385, $N_{\rm Z} = 9$, ][]{zuckerman_aluminumcalcium-rich_2011}.}
    \label{fig:acc_rates}
\end{figure}

The total accretion rate calculated for LSPM\,J0207+3331 is $\log \dot{M} = 9.81$ g s$^{-1}$, placing it among the most actively accreting hydrogen-rich white dwarfs known. Figure~\ref{fig:acc_rates} compares this value with the revised sample from \citet{blouin_no_2022}, where LSPM\,J0207+3331 lies in the upper envelope of the observed distribution.

This high accretion rate reinforces the conclusion of \citet{blouin_no_2022} that there is no significant decline in accretion activity with increasing cooling age between 1 and 8 Gyr, contrary to earlier studies that suggested such a trend \citep[e.g.,][]{hollands_cool_2018,chen_power-law_2019}. For consistency, we also computed the accretion rate using the normalized prescription of \citet{koester_accretion_2009} adopted by \citet{blouin_no_2022}, namely $\dot{M}_{\rm WD} = \dot{M}_{\rm Ca}/0.016$, which assumes bulk Earth abundances for the total composition. Applied to the Ca abundance in LSPM\,J0207+3331, this yields $\log \dot{M} = 9.91$ g s$^{-1}$, consistent with the value obtained from our detailed atmospheric analysis. This agreement confirms that the simplified approach provides a reasonable estimate of the total accretion rate in this case.

The total mass of heavy elements currently residing in the convection zone is $1.22\times10^{22}$~g and represents the minimal mass of the accreted body, assuming the accretion of a single body. 
Given the short sinking timescales of 35,000 to 120,000 years, multiple diffusion cycles could already have taken place, since the lifetime of the disk is likely longer than this \citep[$\sim$10$^6$ years,][]{girven_constraints_2012,farihi_circumstellar_2016,cunningham_horizontal_2021}. Consequently, additional metals may have already sunk below the photosphere, and the inferred accreted mass of $\sim10^{22}$ g should be regarded as a lower limit. 

Using this mass estimate and assuming a typical bulk density of $\sim2.1\,\mathrm{g\,cm^{-3}}$, this corresponds to a spherical body with minimum radius of approximately $225$ km which is comparable to the inferred parent body mass and size from many studies on highly polluted white dwarf systems \citep[e.g.,][]{melis_accretion_2011,klein_discovery_2021,obrien_characterizing_2025}. Still, with $M_{\rm Ca} = 3.12\times10^{20}\,{\rm g}$ in its convection zone, LSPM\,J0207+3331 ranks among the top three white dwarfs with cooling age $>$1\,Gyr in terms of accreted calcium mass, and has the highest accreted mass among hydrogen-rich white dwarfs older than 1 Gyr (see bottom panel of Figure \ref{fig:acc_rates}).

\begin{figure}[h]
    \centering
    \includegraphics[width=\linewidth]{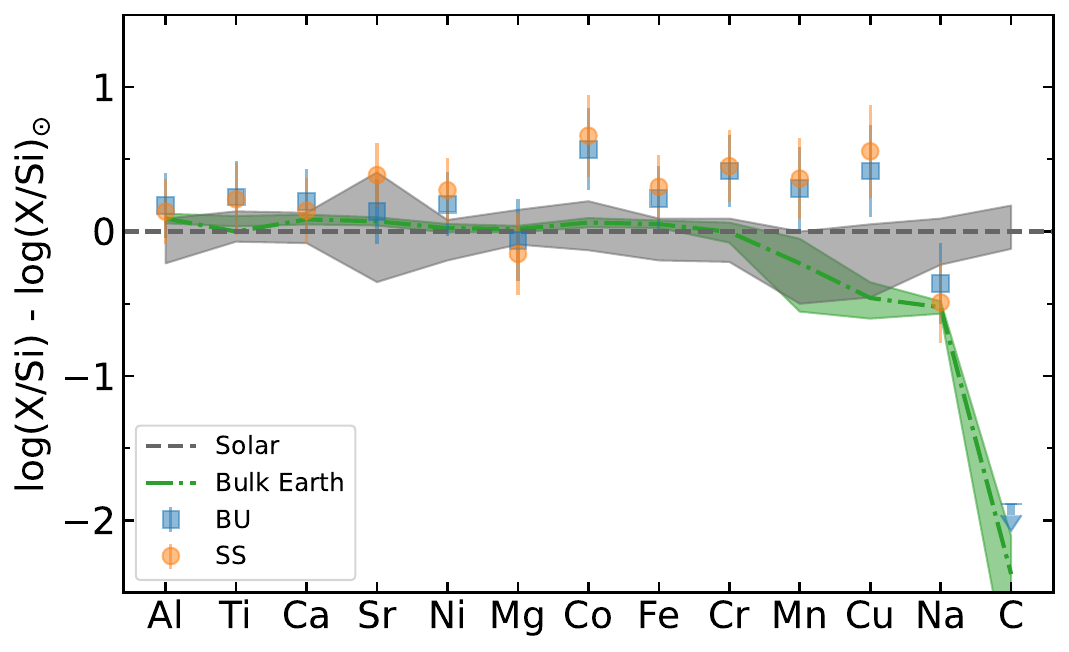}
    \includegraphics[width=\linewidth]{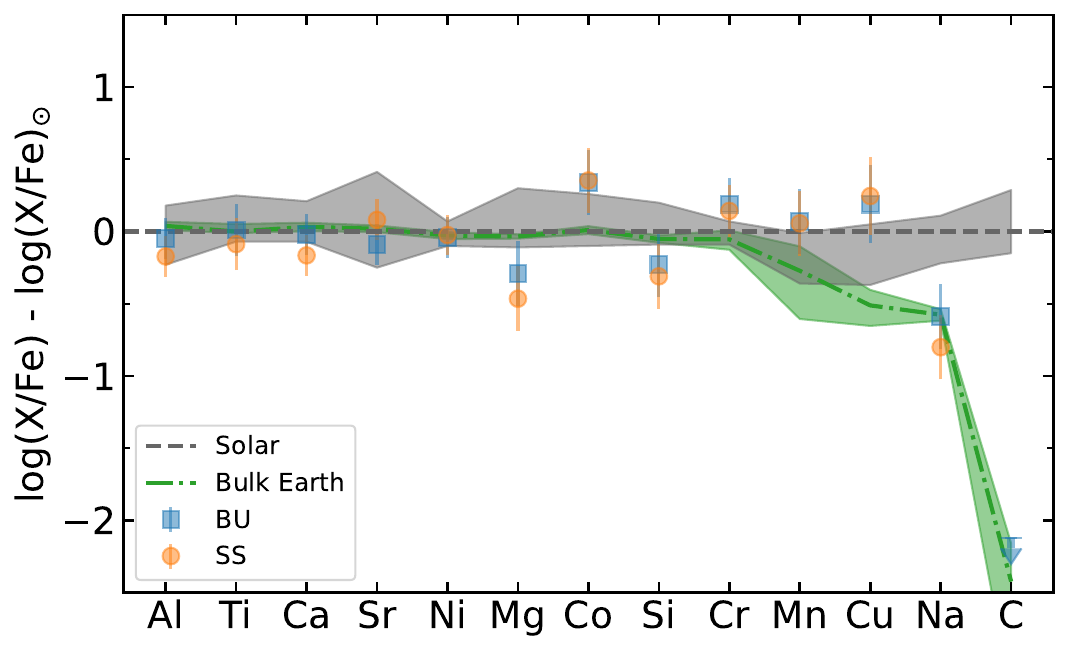}
    \caption{Log abundance ratios relative to Si (top) and Fe (bottom) and normalized to Solar \citep{lodders_solar_2025} for LSPM\,J0207+3331 assuming buildup phase (BU) and steady state (SS). The grey shaded region around solar shows the 95\% range of the abundance ratios of main sequence FG stars from \citet{hinkel_stellar_2014}. Relative abundances for the bulk Earth including uncertainties shown by the green shaded region \citep{wang_elemental_2018} are also plotted for comparison.}
    \label{fig:abundances}
\end{figure}

Figure~\ref{fig:abundances} shows the abundance ratios of the detected elements in the photosphere of LSPM\,J0207+3331 normalized to magnesium (top) and iron (bottom). We compare these measured values to the first two of the three canonical accretion scenarios: (1) the buildup phase, in which the composition reflects that of the accreted body; (2) steady-state, where ongoing accretion is balanced by gravitational settling; and (3) the decreasing phase where accretion has ceased and elements are gradually sinking out of the photosphere. Given the presence of an infrared excess, likely due to a circumstellar dust disk, it is reasonable to assume that the star is currently in the buildup or steady-state phase. In these calculations, we used the metal sinking timescales of \citet{fontaine_timescales_2015}, provided on the Montr\'eal White Dwarf Database \citep{dufour_montreal_2017}, for a pure-hydrogen background. As \citet{fontaine_timescales_2015} did not consider elements heavier than copper, we extended their theoretical framework to compute the timescale of strontium. Furthermore, we verified that considering a hydrogen--helium mixture with $\log ({\rm He/H}) = -1$ has a negligible effect on the sinking timescales, due to the low opacity of helium at the temperature of LSPM\,J0207+3331.

Our adopted timescales do not consider the effect of convective overshoot. This process has been shown to affect the sinking timescales of hydrogen-rich white dwarfs with $11,000\,{\rm K} < T_{\rm{eff}} < 18,000\,\rm K$ \citep{cunningham_convective_2019}, but its impact in cooler white dwarfs such as LSPM\,J0207+3331 is unclear. We verified that adding overshoot over one pressure scale height below the convection zone changes the relative timescales between elements by at most a few percent and is thus unlikely to affect our interpretation of the abundance pattern. We also ignored the effect of thermohaline mixing, which is predicted to be important only at \teff$\,\gtrsim 10,000\,\rm K$ \citep{bauer_polluted_2019}.


Considering these two active accretion scenarios, the abundance ratios of the most refractory elements (Al/Si, Ti/Si, Ca/Si, and Sr/Si) closely match those of the Sun \citep{lodders_solar_2025}, nearby FGK-type main-sequence stars \citep{hinkel_stellar_2014}, and the bulk Earth \citep{wang_elemental_2018}. The Na/Si ratio, and upper limits on C/Si indicate significant depletion in volatile elements, consistent with a dry, rocky body as opposed to volatile-rich material. 

In contrast, the more siderophilic elements (Fe/Si, Ni/Si, Co/Si, Cr/Si and Cu/Si) are all slightly enhanced relative to solar, stellar and bulk Earth values. Even though all individual elements are not enhanced by more than 3\,$\sigma$ relative to solar, their combined enhancement (using Stouffer's method) is significant to $3.1\sigma$ assuming buildup phase and $3.8\sigma$ assuming steady-state phase, suggesting that LSPM\,J0207+3331 could be accreting material enriched with metallic material, perhaps something like a differentiated rocky body's core.
Given the substantial amount of pollution observed in the atmosphere of LSPM\,J0207+3331 and the copious amount of dusty circumstellar material necessary to explain its infrared excess emission (see \citealt{debes_3_2019} and Section \ref{sec:ir-excess}), it would not be unreasonable to assume that the parent body would have been massive enough to have been differentiated and hence feature a metallic core.

Previous examples of white dwarf stars accreting parent bodies with Fe/Mg and Fe/Si enhanced relative to the bulk Earth have also been accompanied by deviant abundances for refractories like Al, Ti, and Ca (e.g., see Figure 6 of \citealt{melis_accretion_2011}). Interpretations for these systems have invoked loss of outer layer (crust+mantle) material during the host star's
post-main sequence evolution (e.g., \citealt{klein_chemical_2010}; \citealt{melis_accretion_2011}). The nearly bulk Earth-like
abundances for LSPM\,J0207+3331 for all measured elements is suggestive
instead of an intact body that has experienced minimal post-main sequence
changes.

Under the assumption of the parent body being a differentiated, massive rocky body, we explore what relative mass fractions its core, mantle, and crust would have had. As a reference, the Earth has a metallic
core that is $\approx$32.5\%
of its total mass \citep{wang_elemental_2018}. We start with generally Earth-like compositions for each layer (a fairly reasonable expectation given Figure \ref{fig:abundances}) and adjust the mass fractions until we arrive at abundance ratios
consistent with what is observed for LSPM\,J0207+3331.
In this manner and considering the buildup phase of accretion, we find that a core mass fraction of $\approx$55\% and crust+mantle mass fraction of $\approx$45\% results specifically in the depressed Mg/Fe and Mg/Si ratios ($\approx$0.62 and $\approx$0.91 by number respectively) while preserving the roughly bulk Earth-like character of other elements. If the steady-state phase of accretion is instead invoked then an even higher core mass fraction is needed. Frustrating a complete assessment of the accuracy of this interpretation is the lack of a measurement or much tighter constraint on oxygen. The O/Fe ratio resulting from our model ($\sim$2 by number) is dramatically lower than the bulk Earth value ($\approx$4.0).

It is interesting to contrast the suggested high core mass fraction for the parent body being accreted by LSPM\,J0207+3331 with recent results in exoplanet structural evolution. Several studies have linked evolving exoplanet properties over the age of the universe
to changing chemical abundance patterns within their host galaxies (e.g., \citealt{cabral23}; \citealt{weeks25}; \citealt{steffen25}). A specific prediction from the results of \citet{steffen25} and \citet{weeks25} is that exoplanet density and hence core mass fractions should generally increase as a function of time as more and more
heavy elements are produced. The planetary system bodies orbiting  LSPM\,J0207+3331 are at least 3\,Gyr old and likely much older (Table \ref{tab:parameters}).
Based on the results of \citet{weeks25} and \citet{steffen25} we should expect a massive, differentiated rocky body in the
LSPM\,J0207+3331 planetary system to have a core mass fraction comparable to that of the Earth or possibly lower. Instead, we find that a reasonable 
interpretation is the presence of a much higher core mass fraction. While it is likely premature to consider our interpretation
for LSPM\,J0207+3331 as final, it is tantalizing as a possible novel means of assessing models for exoplanet formation and evolution on a galactic scale.

Regardless of the exact origin and structure of the parent body being
accreted by LSPM\,J0207+3331, the detection of heavy elements in its atmosphere despite its cooling age exceeding 3 Gyr provides direct evidence that significant accretion episodes can still occur at such advanced stages, supporting the scenario of ongoing delivery of material into the white dwarf's Roche radius proposed by \citet{blouin_no_2022}.

\subsection{\ion{Ca}{2} H and K Core Emission}\label{ssec:ca_core}
\begin{figure}[h]
    \centering
    \includegraphics[width=0.99\linewidth]{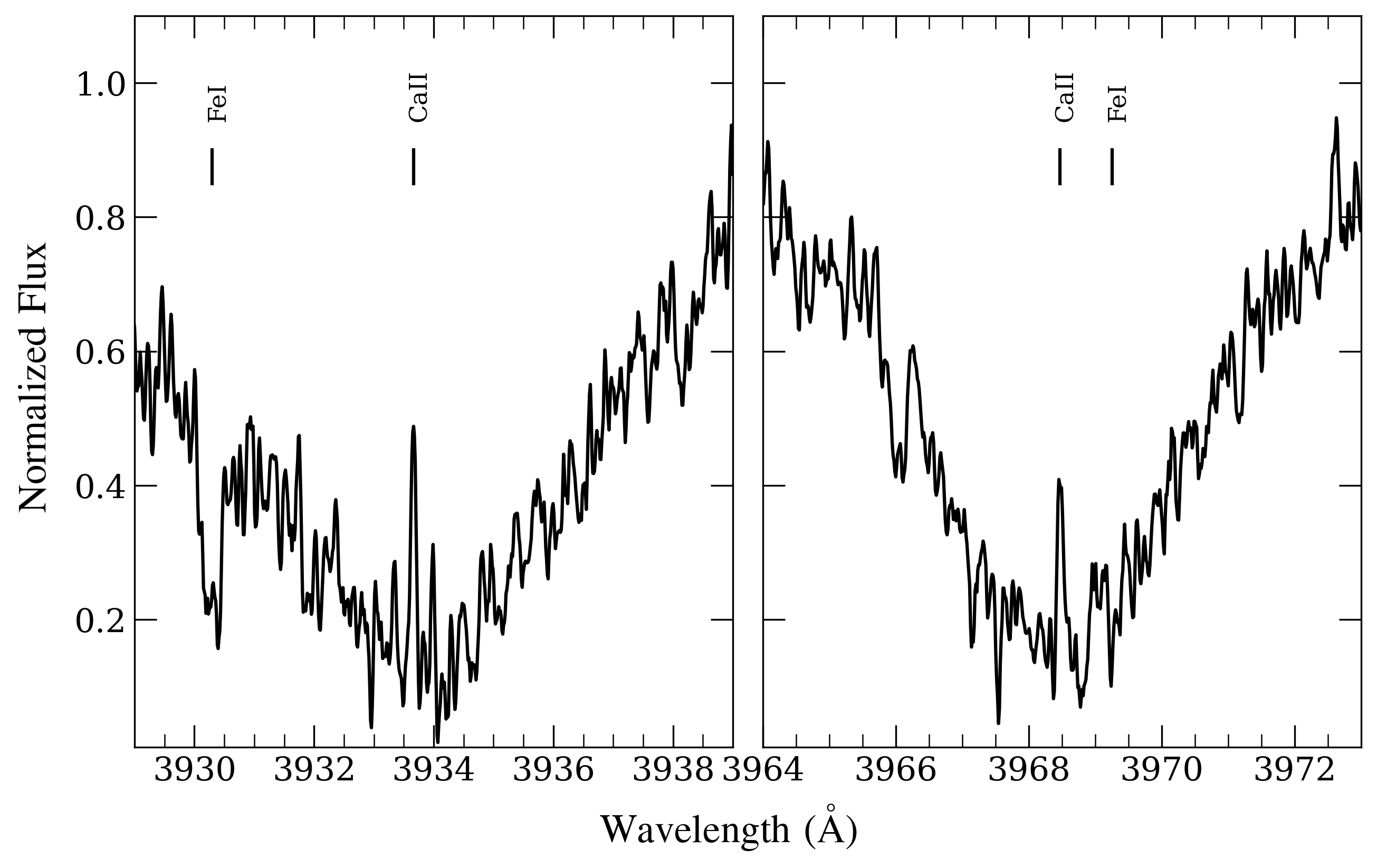}
    \caption{HIRES data of LSPM\,J0207+3331 centered around the \ion{Ca}{2} K (left) and H (right) lines.}
    \label{fig:Ca_core}
\end{figure}

\begin{figure}[h]
    \centering
    \includegraphics[width=0.99\linewidth]{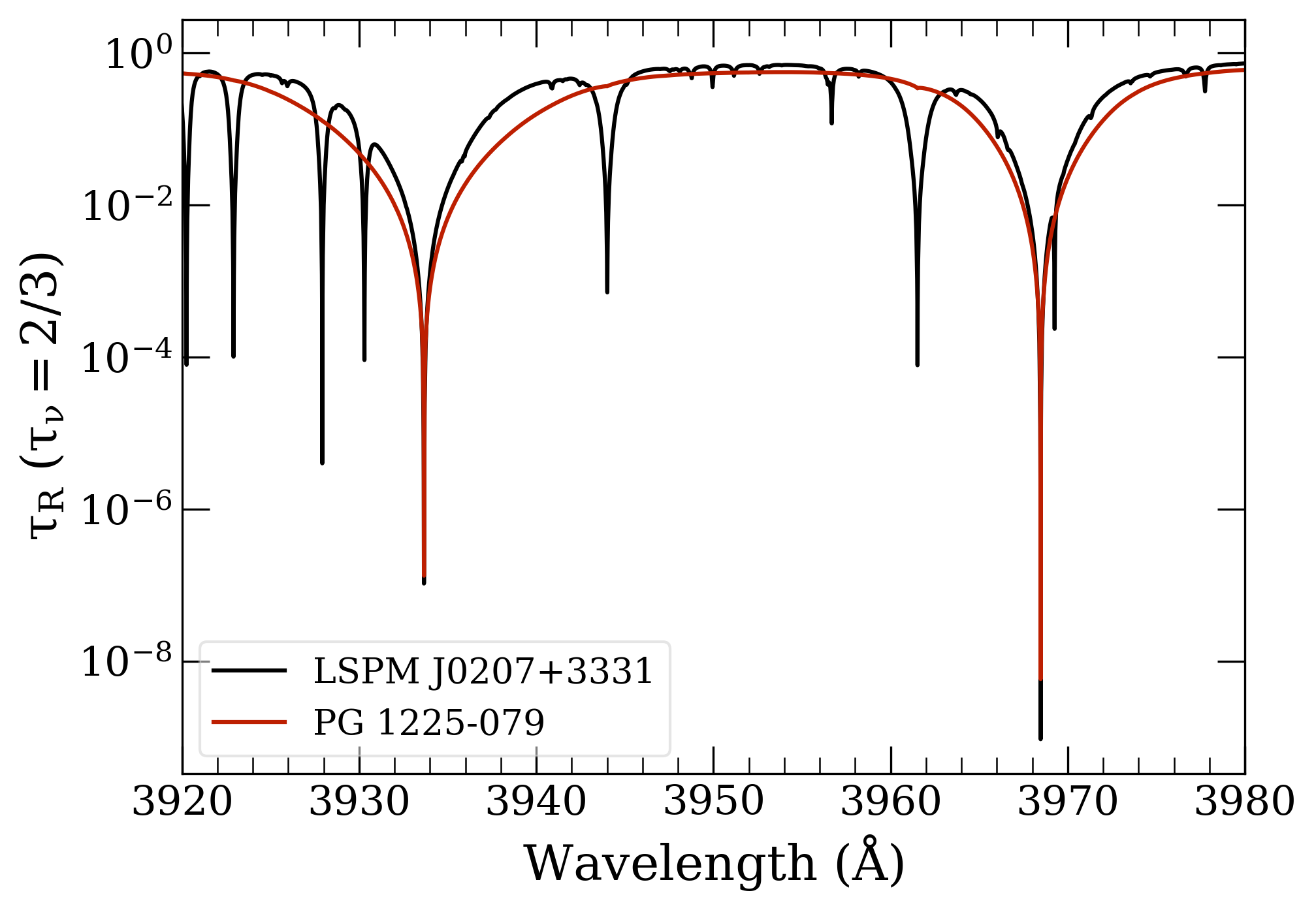}
    \caption{Rosseland mean optical depth at which the monochromatic optical depth reaches $\tau_{\nu} = 2/3$, plotted as a function of wavelength in the vicinity of the \ion{Ca}{2} H \& K lines for LSPM\,J0207+3331 (black) and PG\,1225$-$079 (red). This depth approximately corresponds to the layer from which about 50\% of the emergent photons at a given wavelength escape, making it a useful proxy for line formation depth.}
    \label{fig:tau_ca}
\end{figure}

LSPM\,J0207+3331 exhibits weak but clear signs of \ion{Ca}{2} H \& K core emission (Figure~\ref{fig:Ca_core}, see also Figure~\ref{fig:fit_panel01}), with equivalent widths of 36 and 28 m\AA\ for the K and H components, respectively. Both features are centered at a velocity of approximately 52~km\,s$^{-1}$. To our knowledge, this is only the second reported case of \ion{Ca}{2} core emission in a white dwarf, the first being the helium-rich white dwarf PG1225$-$079 which has 8 metals detected in its atmosphere \citep{klein_rocky_2011}. Similar core emission features are more commonly seen in the \ion{He}{1} $\lambda5876$ line among cool, DB white dwarfs \citep{klein_atmospheric_2020}. In all such cases, the line cores are thought to form in the uppermost atmospheric layers, as illustrated by the depth of formation profile shown in Figure~\ref{fig:tau_ca}.

In main-sequence stars, \ion{Ca}{2} H \& K core emission is a well-established tracer of chromospheric activity, often linked to weak magnetic fields \citep[e.g., ][]{linsky_stellar_2017,cretignier_stellar_2024}. However, our high-resolution HIRES spectrum for LSPM\,J0207+3331 shows no evidence of Zeeman splitting in metal lines, which allows us to place an upper limit of $\sim$10~kG on a possible surface magnetic field.

The presence of \ion{Ca}{2} H \& K core emission, features not reproduced by our LTE photospheric models, suggests that additional physical processes may be at play in or above the upper atmosphere. These could include the presence of a chromosphere, weak magnetic or accretion heating, or deviations from LTE in the outermost layers. Alternatively, the emission may arise from a non-photospheric region altogether, in which case the underlying atmospheric structure remains valid. Regardless of its origin, this second detection of calcium core emission highlights the need for further investigation into the physical mechanisms responsible for this phenomenon in cool, metal-polluted white dwarfs. A detailed exploration of this issue is, however, beyond the scope of the present study.

\section{On the Nature and Origin of the Infrared Excess}
\label{sec:ir-excess}
The origin of the infrared excess around LSPM\,J0207+3331 remains unresolved \citep{bravo_new_2025}. As previously noted, a two-ring model was required by \citet{debes_3_2019} to reproduce the full set of photometric fluxes observed for this star. Studies using the Spitzer Space Telescope have demonstrated the potential for characterizing infrared excesses arising from circumstellar dust and probing its mineralogical composition \citep{reach_dust_2009,xu_dearth_2018}. More recently, using low-resolution spectroscopy with the MIRI instrument on the James Webb Space Telescope (JWST), \citet{farihi_subtle_2025} revealed a striking diversity in 10~$\mu$m silicate emission features among white dwarf debris disks, with several targets exhibiting SEDs similar to that of LSPM\,J0207+3331 (see their Figure A1). In particular, the spectral energy distribution of J0644$-$0352 closely resembles that of LSPM\,J0207+3331, raising the possibility that the apparent WISE $W3$ (11.6~$\mu$m) excess could arise from a strong 10~$\mu$m silicate feature rather than a cold, outer dust ring.

\begin{figure}
    \centering
    \includegraphics[width=\linewidth,trim={0.4cm 0.3cm 0.2cm 0.3cm},clip]{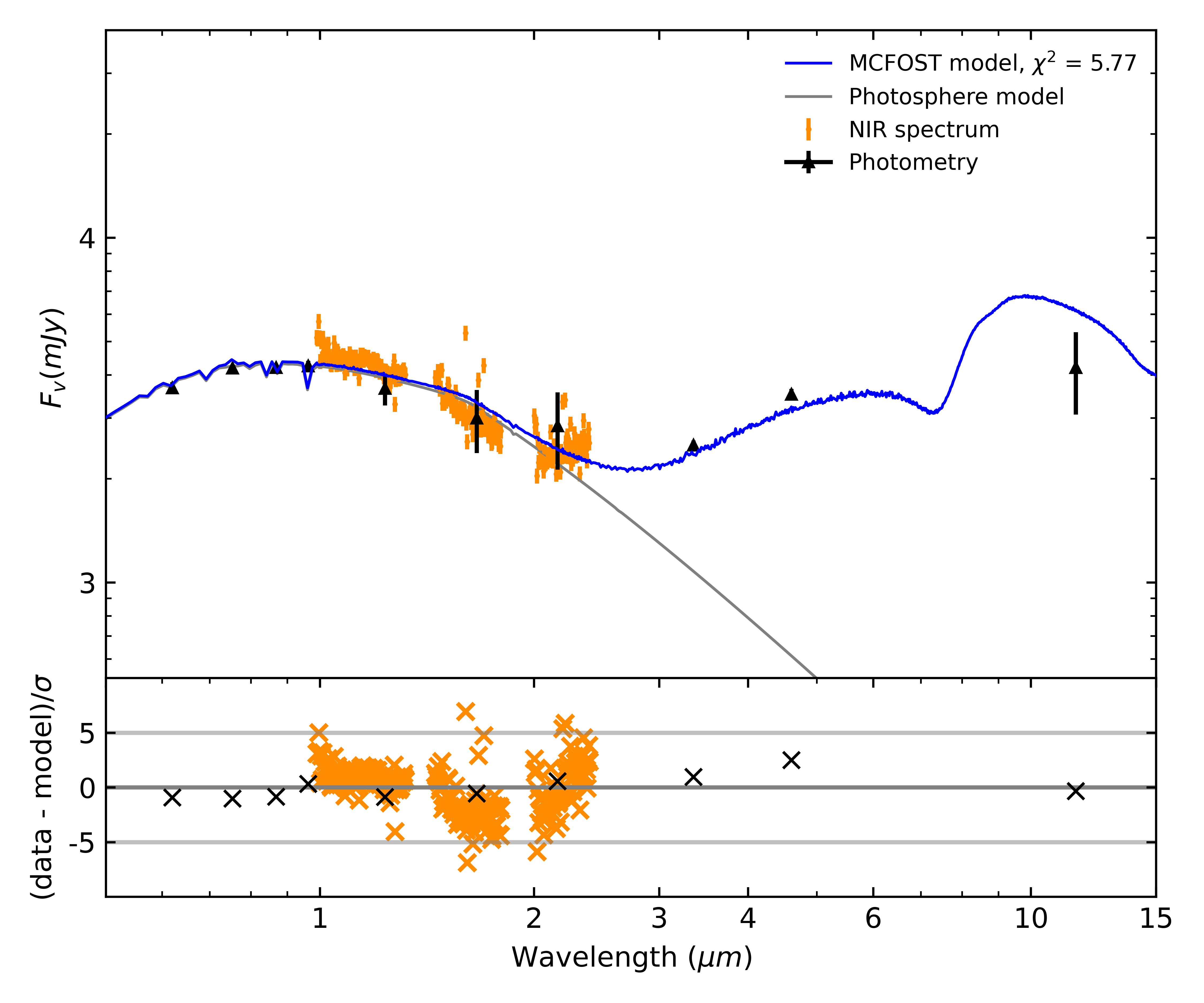}
    \caption{Dust disk model fit (blue) of the photometric (black) and NIRES (orange) data for LSPM\,J0207+3331. The white dwarf SED is plotted in gray.}
    \label{fig:fit_mcfost}
\end{figure}

To test this hypothesis, we modeled the disk using a model with the same properties as in \citet{ballering_geometry_2022}. We fitted the disk inner radius and mass, using a $\chi^2$ minimization method, on the photometry and the NIRES data from \citet{debes_3_2019}. We produced a synthetic spectrum of the dust disk using the radiative transfer code \textsc{mcfost} \citep{pinte_monte_2006,pinte_benchmark_2009}. We find that a model with an inner radius of $R_{\rm in} = 36 \,R_{\rm WD}$, an outer radius of $R_{\rm out} = 54.1\,R_{\rm WD}$ and a total mass of $5.4\times10^{19}{\, \rm g}$ (1.4 -- 1000 microns particles), can reproduce the observations without the need for an additional disk component. The resulting spectral energy distribution is presented in Figure~\ref{fig:fit_mcfost}. Even though the goal of this exercise was to determine if a simple one disk model can describe the infrared excess, the quality of this fit suggests that the 10~$\mu$m silicate feature offers the most plausible explanation.

New JWST observations will be essential to confirm this hypothesis. Determining the dust mineralogy and constraining the total disk mass would also offer valuable insight into the nature of the parent body currently being accreted by LSPM\,J0207+3331.

\section{conclusion}
\label{sec:conclusion}
We report a detailed analysis of the highly polluted, cool, hydrogen-rich white dwarf LSPM\,J0207+3331. Thirteen heavy elements (Na, Mg, Al, Si, Ca, Ti, Cr, Mn, Fe, Co, Ni, Cu, Sr) are detected in its photosphere, the highest number reported to date for a hydrogen-rich white dwarf. This system demonstrates that highly polluted hydrogen-dominated white dwarfs can be just as informative as their helium-rich counterparts in terms of compositional analysis. We also report the fifth detection of strontium in a white dwarf atmosphere, suggesting that this element is more readily detected in cooler white dwarfs, which therefore constitute particularly favorable targets for its observation.

The abundance pattern reveals  volatile-depleted Earth-like material, characterized by enhanced abundances of siderophilic elements (Fe, Ni, Co, Cr, Cu) relative to Mg and Si, suggesting accretion of a differentiated rocky body with a high core mass fraction of about 55\%. This new method for assessing the relative mass fractions of the core, mantle and crust of the accreted material shows potential for testing exoplanet formation and evolution on a galactic scale.   

We detect weak but unambiguous \ion{Ca}{2} H \& K core emission likely originating in the upper atmosphere, the second such case observed in a white dwarf. This feature points to the presence of additional physical processes, such as chromospheric activity, that merit further investigation.

Our analysis demonstrates that high metal pollution significantly alters the atmospheric structure of cool hydrogen-rich white dwarfs. In particular, for white dwarfs with $\log({\rm Ca/H}) \gtrsim -8.0$ and $T_{\rm eff} \lesssim 5000$~K, neglecting heavy elements in the structural calculations introduces systematic biases in photometrically derived parameters, especially in the near-infrared (e.g., 2MASS $JHK_s$).

Finally, the high accretion rate ($\log \dot{M} = 9.81$~g\,s$^{-1}$), combined with the short sinking timescale of strontium ($\sim$35,000 years) relative to the lifetime of the disk and the strong infrared excess, support ongoing accretion rather than a single past event. This implies that the remnant planetary system contains a substantial reservoir of debris capable of sustaining pollution over gigayear timescales--long after the death of the progenitor star.

\begin{acknowledgments}
The authors thank the anonymous referee for their helpful comments to improve the clarity of this manuscript. The authors wish to thank B. Zuckerman for his help obtaining the Keck HIRES data. We acknowledge the support of the Natural Sciences and Engineering Research Council of Canada (NSERC). E. L. and P. D. acknowledge the financial support of the Trottier Family Foundation. This work has made use of the VALD database, operated at Uppsala University, the Institute of Astronomy RAS in Moscow, and the University of Vienna. This work has made use of the Montr\'eal White Dwarf Database. 
C. M.\ acknowledges support from NSF grant SPG-1826583. 
S. Xu and L. K. Rogers are supported by the international Gemini Observatory, a program of NSF NOIRLab, which is managed by the Association of Universities for Research in Astronomy (AURA) under a cooperative agreement with the U.S. National Science Foundation, on behalf of the Gemini partnership of Argentina, Brazil, Canada, Chile, the Republic of Korea, and the United States of America.
Research at Lick Observatory is partially supported by a generous gift from Google.
This paper includes data gathered with the 6.5 meter Magellan Telescopes located at Las Campanas Observatory, Chile.
Some of the data presented herein were obtained at the W.M. Keck Observatory, which is operated as a scientific partnership among the California Institute of Technology, the University of California and NASA. The Observatory was made possible by the generous financial support of the W. M. Keck Foundation. We recognize and acknowledge the very significant cultural role and reverence that the summit of Maunakea has always had within the indigenous Hawaiian community.
\end{acknowledgments}

\appendix

Table \ref{tab:lines} presents lines used for fitting the optical spectrum
of LSPM\,J0207+3331.

\begin{deluxetable*}{cl}[h]
\tablecaption{Lines used for fitting.}\label{tab:lines}
\tablehead{
\colhead{Ion}&\colhead{Air Wavelength (\AA)}
}
\startdata
\ion{Na}{1} & 5889.951, 5895.924\\
\ion{Mg}{1} & 3829.355, 3832.299, 3832.304, 3838.292, 3838.295, 5172.684, 5183.604 \\
\ion{Al}{1}  &  3829.355, 3944.006, 3961.520 \\
\ion{Si}{1} & 3905.523\\
\ion{Ca}{1} &  4226.728 \\
\ion{Ca}{2} &  3706.024, 3736.902, 3933.663, 3968.469\\
\ion{Ti}{2} & 3322.934, 3329.453, 3335.191, 3341.874, 3349.033, 3349.402, 3361.212, 3372.793, 3383.759, 3387.834, 3394.572, \\
& 3685.189, 3685.204, 3759.291, 3761.320\\
\ion{Cr}{1} & 3578.686, 3593.485, 3605.329, 4254.336, 4274.797, 4289.717, 4351.811, 5204.511, 5206.037, 5208.425\\
\ion{Cr}{2} &  3368.041 \\
\ion{Mn}{1} & 3547.790, 3548.019, 3569.490, 3806.711\\
\ion{Mn}{2} & 3441.985, 3460.314, 3474.038, 3474.127, 3482.904, 3488.675\\
\ion{Fe}{1} &  3440.606, 3440.989, 3443.876, 3465.861, 3475.450, 3476.702, 3490.574, 3497.841, 3513.818, 3521.261, 3526.041, \\
& 3526.166, 3554.924, 3558.515, 3565.379, 3570.098, 3570.254, 3581.193, 3585.319, 3585.705, 3586.985, 3606.679, \\
& 3608.859, 3618.768, 3631.463, 3647.843, 3679.913, 3719.930, 3734.864, 3737.132, 3743.362, 3745.561, 3745.899, \\
& 3748.262, 3749.485, 3758.233, 3763.789, 3767.192, 3787.880, 3795.002, 3798.511, 3799.547, 3812.964, 3815.840, \\
& 3820.425, 3824.444, 3825.881, 3827.822, 3834.222, 3840.437, 3841.048, 3849.966, 3856.371, 3859.911, 3865.523, \\
& 3872.501, 3878.018, 3878.573, 3886.282, 3887.048, 3888.513, 3895.656, 3899.707, 3902.945, 3906.479, 3920.258, \\
& 3922.912, 3927.920, 3930.297, 3969.257, 4063.594, 4071.738, 4132.058, 4143.868, 4199.095, 4202.029, 4250.787, \\
& 4271.760, 4294.125, 4307.902, 4325.762, 4383.545, 4404.750, 4415.122, 5167.488, 5227.189, 5269.537 \\
\ion{Co}{1} & 3483.405, 3502.278, 3506.312, 3512.637\\
\ion{Ni}{1} & 3414.760, 3423.704, 3433.554, 3437.275, 3446.255, 3452.885, 3458.456, 3461.649, 3492.954, 3500.846, 3510.332, \\
& 3515.049, 3524.535, 3566.366, 3571.860\\
\ion{Cu}{1} & 3247.537, 3273.954\\
\ion{Sr}{2} & 4077.709, 4215.519\\
\enddata

\end{deluxetable*}

\bibliography{bibliography}{}
\bibliographystyle{aasjournal}


\end{CJK*} 

\end{document}